\documentclass[
a4paper,
twocolumn,
floatfix,
accepted=2025-07-08,
10pt]{quantumarticle}

\pdfoutput=1

\usepackage[numbers]{natbib}
\usepackage{bbm}
\usepackage{amsmath}
\usepackage{amssymb}
\usepackage{amsthm}
\usepackage{amsfonts}
\usepackage{amsmath}
\usepackage{graphicx}
\usepackage{xcolor}
\usepackage[colorlinks=true,linkcolor=teal,citecolor=teal,urlcolor=teal]{hyperref}
\usepackage{mathtools}
\usepackage{booktabs} 
\usepackage{bm}
\usepackage[export]{adjustbox}
\usepackage{multirow}
\usepackage{mleftright}
\usepackage[normalem]{ulem}
\usepackage[ruled,lined]{algorithm2e}
\usepackage{tabularx}
\usepackage{physics}
\usepackage{orcidlink}
\usepackage{mathtools}

\DeclarePairedDelimiter{\bbra}{\langle\!\langle}{|}
\DeclarePairedDelimiter{\kket}{|}{\rangle\!\rangle}

\newcommand{\bbrakket}[2]{\bbra{#1} {#2}\rangle\!\rangle}

\newcommand{\identity}{\mathbb{I}}
\newcommand{\diag}{\operatorname{diag}}
\newcommand{\E}[1]{\mathbb{E}\qty[#1]}
\newcommand{\Var}[1]{\text{Var}\qty[#1]}

\DeclareMathOperator*{\argmin}{arg\,min}
\newcommand{\eg}{\textit{e.g.} }
\newcommand{\ie}{\textit{i.e.} }

\newcommand{\overbar}[1]{\mkern 1.5mu\overline{\mkern-1.5mu #1 \mkern-1.5mu}\mkern 1.5mu}

\newlength\mylen
\newcommand\myinput[1]{%
  \settowidth\mylen{\KwIn{}}%
  \setlength\hangindent{\mylen}%
  \hspace*{\mylen}#1\\}

\SetKw{Continue}{continue}

\newtheoremstyle{sltheorem}
{}                
{}                
{\slshape}        
{}                
{\bfseries}       
{.}               
{ }               
{}                
\theoremstyle{sltheorem}
\newtheorem{theorem}{Theorem}[section]

\theoremstyle{definition}

\begin{document}

\title{Low variance estimations of many observables with tensor networks and informationally-complete measurements}

\author{Stefano Mangini}
\orcid{0000-0002-0056-0660}
\email{stefano.mangini@algorithmiq.fi}
\affiliation{Algorithmiq Ltd, Kanavakatu 3C 00160 Helsinki, Finland.}
\affiliation{QTF Centre of Excellence, Department of Physics, University of Helsinki, P.O. Box 43, FI-00014 Helsinki, Finland.}

\author{Daniel Cavalcanti}
\orcid{0000-0002-2704-3049}
\email{daniel@algorithmiq.fi}
\affiliation{Algorithmiq Ltd, Kanavakatu 3C 00160 Helsinki, Finland.}

\begin{abstract}
Accurately estimating the properties of quantum systems is a central challenge in quantum computing and quantum information. We propose a method to obtain unbiased estimators of multiple observables with low statistical error by post-processing informationally complete measurements using tensor networks. Compared to other observable estimation protocols based on classical shadows and measurement frames, our approach offers several advantages: (i) it can be optimized to provide lower statistical error, resulting in a reduced measurement budget to achieve a specified estimation precision; ({ii}) it scales to a large number of qubits due to the tensor network structure; ({iii}) it can be applied to any measurement protocol with measurement operators that have an efficient tensor-network representation. We benchmark the method through various numerical examples, including spin and chemical systems, and show that our method can provide statistical error that are orders of magnitude lower than the ones given by classical shadows.
\end{abstract}

\maketitle

\section{Introduction}
Extracting and interpreting information from quantum systems is a fundamental task in quantum information processing. A key approach is quantum state tomography (QST), which aims to reconstruct a quantum state from a complete set of measurement data that uniquely characterizes it. However, since the number of parameters required to describe a quantum state grows exponentially with the system size, standard tomography is limited to either very small systems or to states that admit a compact representation~\cite{EfficientQSTCramer2010, QST-GM-Carrasquilla2019, QST-LPDO-Guo2024}.

In many practical scenarios, one is not interested in recovering the full state but rather to estimate expectation values of some observables on it. In this cases, tomographically complete measurements can be used to directly estimate such quantities circumventing a costly characterization of the full state~\cite{Perinotti2007optimalestimationensembleaverages, OptimalProcessingDariano2007}. An advantage of this protocol lies in its ``measure first, ask later" approach: initially, one runs the quantum experiments to collect informationally-complete measurement data on the given state, and later post-process such data to estimate the expectation value of any observable of interest.

This idea has recently been revisited and extended under the name of \emph{shadow estimation}~\cite{HuangShadows2020}, where information completeness is achieved through randomized measurements. Specifically, performance guarantees have been established for certain classes of informationally complete measurements, demonstrating the efficiency of this approach in various settings. Despite significant theoretical and experimental progress, shadow estimation still faces two main challenges. The first is the difficulty of inverting the measurement channel, an essential step in the post-processing of expectation values. Unless approximate inversions are employed~\cite{Cioli2024approximateinversemeasurementchannel}, this restricts shadow estimation to specific types of measurements~\cite{HuangShadows2020, KohNoisyShadows2020, FermionicShadowTomography2021, Hu2021ShallowShadow, ChenRobustShadow2021, Bertoni2023shallow, Akhtar2023ShallowShadowTN, OnoratiNoisyShadows2024, Farias2024RobustShallowShadow}, thus limiting the exploration of a wider range of measurement schemes that could further improve the estimation efficiency. The second challenge is that, in many cases, the inverse of the measurement channel does not result in the most efficient estimator in terms of measurement overhead. Building on the equivalence between classical shadows and quantum measurement frames~\cite{ShadowTomographyDualInnocenti2023, DArianoICMeasurements2004, ScottTightICPOVM2006, ZhouOCPOVM2014}, this issue has been recently investigated in Refs.~\cite{Malmi2024EnhancedEstimation, CaprottiDualOptimisation, Fischer2024DualOptimization}, but the proposed techniques don't scale to large system sizes and can only take into account local measurements and post-processing. These barriers are a limiting factor in finding measurements and post-processing methods that can reduce the measurement overhead of randomized measurement schemes.

In this paper, we present a different way of post-processing the data coming from informationally complete measurements that overcomes the two barriers described above. Our method circumvents the channel inversion needed in the shadow protocol and provides unbiased estimators through an efficient parameterization based on tensor networks. In particular, it can be applied to any estimation procedure for which the observables and measurement operators admit an efficient representation in terms of tensor networks ---this also includes measurement strategies based on \textit{shallow shadows} using possibly noisy random shallow circuits~\cite{Hu2021ShallowShadow, ChenRobustShadow2021, Bertoni2023shallow, Akhtar2023ShallowShadowTN, OnoratiNoisyShadows2024, Farias2024RobustShallowShadow}. 

Importantly, when informationally \textit{overcomplete} measurements are used (which is the case of shadow estimation based on Clifford measurements~\cite{HuangShadows2020}), the method provides optimised unbiased estimators with low variance, outperforming classical shadows in terms of sample efficiency, that is the number of measurement shots needed to achieve a certain error. As we show numerically on several spin and chemical examples up to $n=22$ qubits, for reasonably low bond dimensions our tensor-network estimator can decrease by orders of magnitude the statistical errors associated with the estimations, and even reach optimal variance estimations.

\begin{figure*}[!ht]
    \centering
    \includegraphics[width=\textwidth]{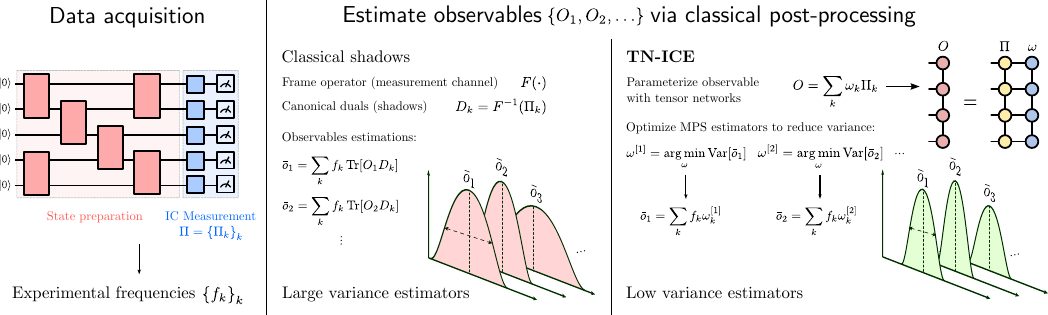}
    \caption{Schematic representation of the proposed method. First, one collects measurement data by measuring the system with an informationally (over-)complete measurement. This can be implemented either by means of randomized measurements schemes (such as in the classical shadow protocol), or with more general IC measurement implementations. Thanks to informational completeness, the same measurement dataset can be later used to estimate many observables in post-processing. The classical shadows protocol uses canonical dual frame operators to construct the observable estimation. Depending on the case, such state agnostic post-processing can however result in estimators with a large statistical error (variance). We instead propose TN-ICE, a new method that uses tensor network to post-process the data by finding a decomposition of the observable in terms of the measurement operators so that the resulting estimator has a low statistical error. This is done by parameterizing the expansion coefficients $\qty{\omega_k}$ as a Matrix Product State (MPS) and optimizing them to find low statistical error estimators.}
    \label{fig:main-image}
\end{figure*}

We summarize the main idea of the manuscript in Fig.~\ref{fig:main-image}. The rest of the manuscript is organised as follows:
\begin{itemize}
    \item In Sec.~\ref{sec:ic-measurement} we review informationally-complete measurements for multiple-observable estimation tasks, which is at the basis of classical shadows and also our method.
    \item In Sec.~\ref{sec:our-contribution} we summarize the main contributions of this work.
    \item In Sec.~\ref{sec:dual-frames} we revisit the post-processing method used in shadow estimation within the broader framework of measurement frames. This section aims to clarify the challenges associated with the existing estimation methods and prepare the reader to understand how our approach works and how it extends beyond shadow estimation. 
    \item In Sec.~\ref{sec:tn-estimation} we present our tensor network-based estimation protocol. We discuss how it can be used to provide reliable estimators with low variance, provide an explicit DMRG-like procedure to optimize it, and discuss the methods' performances and error guarantees.
    \item In Sec.~\ref{sec:numerics} we provide numerical results, including the estimation of a global observable on a GHZ state and the energy estimation of several molecular systems. We also analyse the method’s performance under finite statistics and propose strategies to avoid overfitting.
    \item Finally, in Sec.~\ref{sec:final-remarks} we provide concluding remarks, including issues that we believe are worth exploring in light of the results presented in this paper.
\end{itemize}

\section{Simultaneous estimation of many observables with informationally complete measurements} 
\label{sec:ic-measurement}
A quantum measurement with $r$ possible outcomes is described by a positive operator value measure (POVM) which consists of a set of positive operators $\{\Pi_k\}_{k=1}^r$ that sum to the identity, \ie $\Pi_k \geq 0~\forall~r$ and $\sum_k \Pi_k = \identity$. The measurement operators $\Pi_k$ are usually called measurement or POVM effects and provide the probability $p_k$ of obtaining result $k$ upon measuring a system in a state $\rho$ through $p_k=\Tr[\Pi_k \rho]$. A measurement is called informationally complete (IC) if its effects span the operator space, meaning that any observable $O$ can be written as~\cite{DArianoICMeasurements2004} 
\begin{equation}
\label{eq:operator-decomposition}
O = \sum_{k=1}^r \omega_k \, \Pi_k\,,
\end{equation} 
with $\omega_k \in \mathbb{R}$ being some reconstruction coefficients specific to the observable $O$. Informationally complete measurements have $r \geq d^2$ effects, with $d^2$ of them being linearly independent, where $d$ is the dimension of the Hilbert space (for example, $d=2^n$ for a system of $n$ qubits).

The operator reconstruction formula~\eqref{eq:operator-decomposition} can be used to compute expectation values by averaging the reconstruction coefficients over the measurement statistics as follows
\begin{equation}
    \label{eq:mean-estimator}
    \expval{O} = \Tr[O\rho] = \sum_{k} \omega_k \Tr[\Pi_k \rho] = \sum_k p_k\, \omega_k\,.
\end{equation}
Thus, one can obtain the expectation value of any observable by computing the average of the reconstruction coefficients $\{\omega_k\}$, considered as possible realizations of a discrete random variable $\omega$ distributed according to the probabilities $\{p_k\}$.

In a real experiment, we only have access to a finite number of measurement shots $S$, in which case the expectation value $\expval{O}$ can be inferred through the unbiased estimator given by the sample mean
\begin{equation}
    \label{eq:unbiased-estimator}
    \overbar{\omega} = \sum_k f_k \, \omega_k = \frac{1}{S}\sum_{s=1}^{S}\omega_{k_s}\,,
\end{equation}
where $f_k$ are the experimental frequencies of measuring outcome $k$, and $\omega_{k_s}$ corresponds to the reconstruction coefficient for outcome $k$ obtained in the $s$-th experimental shot. Indeed, one can easily check that this estimator is unbiased, that is $\mathbb{E}[\overbar{\omega}]=\expval{O}$. The standard error of the sample mean is defined by
\begin{equation}
    \label{eq:std}
   \sigma[\overbar{\omega}] = \sqrt{\frac{\text{Var}[\omega]}{S}},
\end{equation}
where
\begin{equation}
    \label{eq:estimator-variance}
   \text{Var}[\omega] \coloneqq \sum_{k=1}^r f_k\, \omega_k^2 - \qty(\sum_{k=1}^r f_k \omega_k)^2,
\end{equation}
is the sample variance, which is itself as an estimate of the true variance of the underlying distribution. Assuming a large number of samples, the central limit theorem implies that the sample mean converges to the true mean $\overbar{\omega}\rightarrow \expval{O}$ as $\sqrt{\Var{\omega}/S}$. Since the true variance of the distribution is unknown, we replace it with the empirical sample variance.

\section{Summary of contributions}
\label{sec:our-contribution}
Our primary goal is to develop efficient, unbiased estimators based on informationally complete measurements, allowing for the simultaneous estimation of multiple observables from the same dataset. The term ``efficient" has two meanings in this context. First, the estimators should yield low-variance results, thereby minimizing the measurement overhead required to achieve a given estimation accuracy. Second, they should be computationally efficient, scaling favourably with the number of qubits.

To address the first criterion, we formulate the following optimization problem:
\begin{align}\label{eq:min-variance}
    &\min_{\{\omega_k\}}~\Var{\omega} = \sum_{k=1}^r f_k \omega_k^2 - \qty(\sum_{k=1}^r f_k \omega_k)^2, \\
    &\text{subject to} \quad O = \sum_{k=1}^r \omega_k \Pi_k.
\end{align}
In other words, given an observable $O$ and measurement data $\{f_k\}$, our goal is to minimize the variance of the estimator while ensuring that the coefficients $\{\omega_k\}$ define an unbiased estimator for $O$. It is important to highlight the \emph{multi-observable} nature of our approach: using the same dataset $\{f_k\}$, one can solve the optimization problem for different observables, yielding low-variance estimators for each.

To address the second criterion, we propose parameterizing the relevant quantities in this optimization as tensor networks with controlled bond dimensions. The efficiency of the method thus depends on how compactly (\ie with what bond dimension) the measurement operators and observables can be represented within a tensor network framework. Notably, many commonly used measurements, such as local measurements or those derived from shallow circuits, admit efficient tensor network representations. Similarly, many observables of interest ---such as correlation functions, overlaps with matrix-product states, and energy estimations for many spin chains and molecular systems--- can also be efficiently represented as tensor networks. Finally, the bond dimension used to parametrise the observable's expansion coefficients $\{\omega_k\}$ serves as a tunable parameter, balancing the classical computational cost of the method against the achievable minimum variance.

We will refer to the estimator obtained through this method as TN-ICE, from Tensor Network Informationally Complete Estimator.

\section{Relation with measurement frames and classical shadows}
\label{sec:dual-frames}
The standard approach to obtain unbiased estimators of observables from IC measurement data is through the use of \textit{dual frames}~\cite{CasazzaFiniteFrames2013, KrahmerSparsityDualFrames2013, ShadowTomographyDualInnocenti2023, DArianoICMeasurements2004, ZhouOCPOVM2014}. Given an IC-POVM with effects $\{\Pi_k\}$, the set of operators $\{D_k\}$ forms a dual frame to $\{\Pi_k\}$ if the following decomposition formulas hold \textit{for every operator} $O$ 
\begin{equation}
    \label{eq:can-dual-obs-decomp}
    O = \sum_k \Tr[O\,D_k]\, \Pi_k = \sum_k \Tr[O\,\Pi_k]\, D_k\,.
\end{equation}
This means that the reconstruction coefficients in Eq.~\eqref{eq:operator-decomposition} are of the form $\omega_k = \Tr[O D_k]$, where $\{D_k\}$ are the dual frame (or dual effects) of $\{\Pi_k\}$. In the case that the IC-POVM has exactly $d^2$ measurement effects it is called a minimal IC-POVM, and its dual effects are uniquely defined. On the other hand, if the IC measurement contains more than $d^2$ effects it is called an informationally over-complete (OC) POVM, and some of its effects are not linearly independent from the rest. Importantly, in this case the choice of the duals effects is not unique~\cite{ShadowTomographyDualInnocenti2023, DArianoICMeasurements2004, ZhouOCPOVM2014}.

A particular choice of dual frame is given by the canonical duals, obtained by~\cite{ZhouOCPOVM2014, ShadowTomographyDualInnocenti2023}\footnote{For simplicity we are ignoring some subtleties related to the definition of canonical duals used in the context of quantum tomography~\cite{ZhouOCPOVM2014}, or in the context of general frame theory for linear algebra~\cite{CasazzaFiniteFrames2013}. We refer to~\cite{ShadowTomographyDualInnocenti2023} for an extended discussion on the topic.}.
\begin{equation}
    \label{eq:canonical-dual}
    D_k = \frac{1}{\Tr[\Pi_k]} F^{-1}\qty(\Pi_k)\,,
\end{equation} 
where $F$ is a linear map, the so-called \textit{frame operator}, defined as 
\begin{equation}
    \label{eq:canonical-dual-map}
    F(\bm{\cdot}) \coloneqq \sum_k \frac{1}{\Tr[\Pi_k]} \Tr[\, \bm{\cdot}\, \Pi_k]\,\Pi_k\,.
\end{equation}

As discussed in~\cite{ShadowTomographyDualInnocenti2023}, the shadow estimation protocol~\cite{HuangShadows2020} is equivalent to an estimation method based on informationally complete measurements, where canonical dual frames are used to compute unbiased estimators~\cite{DArianoICMeasurements2004, ScottTightICPOVM2006, Perinotti2007optimalestimationensembleaverages, ZhouOCPOVM2014}. Indeed, classical shadows are constructed by applying a random unitary gate to the quantum state, followed by a measurement in the computational basis. Viewed as a single step, this process effectively implements an IC-POVM via a randomized measurements. Such classical shadows are, in fact, the canonical duals to the POVM effects, see Eq.~\eqref{eq:canonical-dual}. Because of this, we will refer to these methods interchangeably as classical shadows or canonical estimations. We refer the interested reader to Ref.~\cite{ShadowTomographyDualInnocenti2023} and Appendix~\ref{app:dual-frames-classical-shadows} for more details on the topic, where we also provide an explicit demonstration of such equivalence for the specific case of local Pauli measurements.

One of the main limitations of canonical (or shadow) estimation arises from the need to compute the dual effects of a general measurement by inverting the exponentially large (in the number of qubits $n$) frame operator ---or measurement channel, in shadow terminology--- in Eq.~\eqref{eq:canonical-dual-map}. Due to this challenge, the use of such estimators has been largely restricted to specific measurement strategies, namely: (\textit{i}) local measurements acting individually on each qubit, these admit a tensor product structure for the inversion map and hence the duals~\cite{HuangShadows2020, Paini2021estimating}; (\textit{ii}) global Clifford measurements, which allow for an explicit and classically tractable computation of the same quantities~\cite{HuangShadows2020, Paini2021estimating}, but they cannot be efficiently implemented on hardware; and  
(\textit{iii}) schemes based on specific shallow measurement circuits~\cite{Hu2021ShallowShadow, Arienzo2022ShallowShadow, Bertoni2023shallow, Akhtar2023ShallowShadowTN, Hu2024RobustShallowShadow, Farias2024RobustShallowShadow}, whose application is however hindered by the presence of noisy gates. 

Additionally, since the choice of dual operators is not unique for overcomplete POVMs (OC-POVMs), one can think of looking for the dual frame that minimize the variance of the resulting estimator. These \textit{optimal} duals have been characterized in refs.~\cite{ShadowTomographyDualInnocenti2023, DArianoICMeasurements2004}, where an an explicit expression for constructing them is provided. However, computing such duals requires prior knowledge of the quantum state being measured, which is generally inaccessible in practical scenarios where full state tomography is infeasible. To overcome this limitation, recently some methods have been proposed to optimize the duals for achieving low-variance estimators without requiring explicit knowledge of the state~\cite{Malmi2024EnhancedEstimation, Fischer2024DualOptimization, CaprottiDualOptimisation}. These approaches however do not scale efficiently with the system size and are mainly limited to local measurements and duals.

\section{\label{sec:tn-estimation} TN-ICE: efficient low-variance estimations with overcomplete measurements and tensor networks}
As mentioned in Sec.~\ref{sec:our-contribution}, the core idea behind TN-ICE is to directly look for the reconstruction coefficients $\smash{\{\omega_k\}}_k$ for which the unbiased estimator in Eq.~\eqref{eq:unbiased-estimator} attains the lowest variance. Since the number of coefficients $r$ grows exponentially with the number of qubits $n$, $r \geq 4^n$, an efficient classical representation is therefore necessary. Here we show how this can be achieved by representing the expansion coefficients $\smash{\{\omega_k\}_k}$ as a Matrix Product State (MPS)~\cite{Schollwok2011DMRGMPS} with a given bond dimension $\chi$. Crucially, this not only allows for an efficient classical description of the coefficients but also introduces controllable classical correlations in the post-processing of the measurement data. 

Let us rewrite the unbiased estimator~\eqref{eq:unbiased-estimator} and the estimator variance~\eqref{eq:estimator-variance} in vectorized notation (see Appendix~\ref{app:vectorisation}), which is more convenient to interpret these operations as tensor networks contractions. Let $\Pi$ denote the matrix obtained by stacking together the vectorized effects as
\begin{equation}
    \Pi = 
    \begin{bmatrix}
    \vert & \vert & & \vert \\
    \kket{\Pi_1} & \kket{\Pi_2} & \cdots & \kket{\Pi_r}\\
    \vert & \vert & & \vert
    \end{bmatrix} \in \mathbb{C}^{d^{2}} \times \mathbb{C}^r\,,
\end{equation}
and let $\smash{\kket{\omega} = \begin{bmatrix} \omega_1, \ldots, \omega_r \end{bmatrix} \in \mathbb{R}^r}$ denote the vector of the reconstruction coefficients, and $\kket{O} \in \mathbb{C}^{d^{2}}$ a vectorized representation of the observable $O$. Notice that if one chooses the Pauli basis to perform the vectorization operation, then both $\Pi$ and $\kket{O}$ have only real entries since both are hermitian operators, $\Pi_k = \Pi_k^\dagger$ and $O = O^\dagger$.

With this notation we can concisely express the decomposition formula~\eqref{eq:unbiased-estimator} and the estimator variance~\eqref{eq:estimator-variance} as
\begin{subequations}
\label{eq:matrix-notation-estimator}
\begin{align}
    \kket{O} & = \Pi \kket{\omega} \label{eq:matrix-notation-decomposition} \\
    \Var{\omega} & = \bbra{\omega} P \kket{\omega} - \bbrakket{p}{\omega}^2 \label{eq:matrix-notation-variance}\,,
\end{align}
\end{subequations}
where $\kket{p} = [p_1, \ldots, p_r]$ with $p_k = \bbrakket{\rho}{\Pi_k} = \Tr[\rho \Pi_k]$ is the vector of outcome probabilities, and $P$ is a diagonal matrix defined as $P_{ij} = p_i \, \delta_{ij}$. 

The expectation value $\E{\omega} = \bbrakket{p}{\omega}$ term doesn't actually depend on the reconstruction coefficients, since $\kket{p} = \Pi^\dagger \kket{\rho}$, and thus $\bbrakket{p}{\omega} = \bbra{\rho} \Pi \kket{\omega}$ = $\bbrakket{\rho}{O} = \Tr[O\rho]= \expval{O}$. Note that this is only true in the case of infinite statistics where one has access to the true probabilities, whereas it is only approximately true in the case of finite statistics using frequencies.

We now show how to express Eqs.~\eqref{eq:matrix-notation-estimator} as contractions between structured tensor networks. The main idea is also to represent the effect matrix $\Pi$ in terms of a Matrix Product Operator (MPO), the vectorized observable $\kket{O}$ as a Matrix Product State (MPS), and that the reconstruction coefficients $\kket{\omega}$ also as an MPS. In practice, we will require these representations to be efficient, i.e. to use low bond dimension, so that all contractions can be performed efficiently. This is the case of observables that can be written as a limited sum of Pauli strings (\eg local Hamiltonians, single Pauli strings, magnetisation) and measurements performed through shallows circuits (\eg local measurements, projections over efficiently representable MPS states).

Assuming such efficient tensor network representations are available, then the reconstruction formula~\eqref{eq:matrix-notation-decomposition} can be diagrammatically expressed as a tensor network contraction as
\begin{equation}
\label{eq:tn-diagaram-decomposition}
    \kket{O} = \Pi \, \kket{\omega} \quad \longleftrightarrow \quad \includegraphics[width=0.3\columnwidth, valign=c]{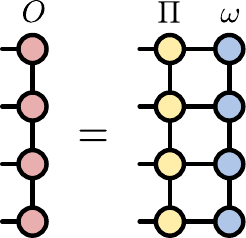} \quad\,,
\end{equation}
where each open leg on the left-hand side of the diagram has dimension $d^2$, and the connected bonds represent summation over the local indices $(k_1, k_2, \ldots, k_n)$ each of size $k_i \in [1, s]$ with $s^n = r$, obtained by expanding the global multi index $k \in [1, r]$ in terms of the local sites.

Similarly, one can also represent the second moment in the estimator variance~\eqref{eq:matrix-notation-variance} in terms of a tensor network contraction as
\begin{equation}
\label{eq:tn-diagaram-variance}
\E{\omega^2} = \bbra{\omega} P \kket{\omega} \quad \longleftrightarrow \quad \includegraphics[width=0.3\columnwidth, valign=c]{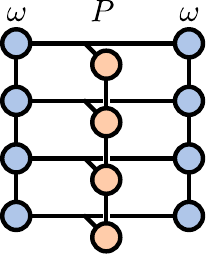}\quad
\end{equation}
where $\omega$ is an MPS as before, and $P$ is an MPO representation of the diagonal probability matrix introduced in Eq.~\eqref{eq:matrix-notation-variance}. Note that while we have represented the probabilities ---or frequencies, in the case of finite statistics--- $P$ as an MPO, this is not a strict requirement as in fact, any efficient classical tensor network representation of the measured outcomes suffices.

The first task we want to solve is to find an MPS $\kket{\omega}$ which provides an unbiased estimation of the observable~\eqref{eq:tn-diagaram-decomposition}, which corresponds to solving the optimization problem
\begin{equation}
    \label{eq:penalty}
     \min_{\kket{\omega}} \big\|\kket{O} - \Pi\, \kket{\omega}\big\|_2^2\,.
\end{equation}
If the solution of this problem is zero, this means that we have found a set of coefficients $\{\omega_k\}_k$ such that \eqref{eq:operator-decomposition} is satisfied and the expression~\eqref{eq:unbiased-estimator} is an unbiased estimator. 

As we have mentioned before, in the case of OC-POVMs (\ie those whose effects form an over-complete basis for the operator space), the decomposition~\eqref{eq:operator-decomposition} is not unique. In this case, it is desirable to look for a set of coefficients that not only provides an unbiased estimator but also attains a low estimation variance~\eqref{eq:tn-diagaram-variance}. This corresponds to solving the following constrained optimization problem
\begin{equation}
\label{eq:constrained_opt}
\begin{aligned}
    \kket{\omega^*} & = \argmin_\omega \Var{\omega | \Pi, O, \rho} = \argmin_{\kket{\omega}} \, \bbra{\omega} P \kket{\omega} \\
    & \quad \text{with } \kket{\omega} \text{ such that } \kket{O} = \Pi\, \kket{\omega}
\end{aligned}\,,
\end{equation}
where one can neglect the first moment term $\mathbb{E}[\omega]$ in the minimization since $\text{Var}[\omega] \leq \mathbb{E}[\omega^2] = \bbra{\omega} P \kket{\omega}$, and it is thus sufficient and more practical to consider minimization of the second moment term only. 

The constrained optimization problem~\eqref{eq:constrained_opt} can be relaxed to the equivalent problem of minimizing the penalty-regularized cost function
\begin{equation}
    \label{eq:cost-function}
    L(\omega) = (1-\lambda)\, \bbra{\omega} P \kket{\omega} + \lambda\, \big\|\kket{O} - \Pi\, \kket{\omega}\big\|_2^2\,.
\end{equation}
which consists of a term proportional to the second moment of the estimator (\ie the first term in the variance \eqref{eq:matrix-notation-variance}), and a second term can be seen as a penalty term forcing~\eqref{eq:operator-decomposition} to hold. The hyperparameter $\lambda \in \mathbb{R}$ weights the importance of the second moment and the penalty term in the cost function, and can be tuned so that the penalty term results in a small value (we will discuss in Sec.~\ref{sec:unbiased-estimation} the impact of the penalty term to the final estimation). The norm of the vectorized operator $\smash{\norm{\kket{A}}_2^2} = \bbrakket{A}{A}$ is equal to the Frobenius norm (or 2-norm) of the operator itself $\smash{\norm{A}_2^2 \coloneqq \Tr[A^\dagger A]}$.

Expanding the norm term and neglecting the constant term $\norm{O}_2$, one can rewrite the cost function as 
\begin{equation}
\label{eq:tn-cost-fun}
\begin{aligned}
    L(\omega) = (& 1 - \lambda)\, \bbra{\omega} P \kket{\omega} \\ 
    & + \lambda\, \qty(\bbra{\omega} \Pi^\dagger \Pi \kket{\omega} - 2\text{Re}\,  \bbra{\omega} \Pi^\dagger \kket{O})\,,
\end{aligned}
\end{equation}
which, using the decompositions~\eqref{eq:tn-diagaram-decomposition} and~\eqref{eq:tn-diagaram-variance}, can be represented in tensor notation as
\begin{equation*}
    \includegraphics[width=\columnwidth, valign=c]{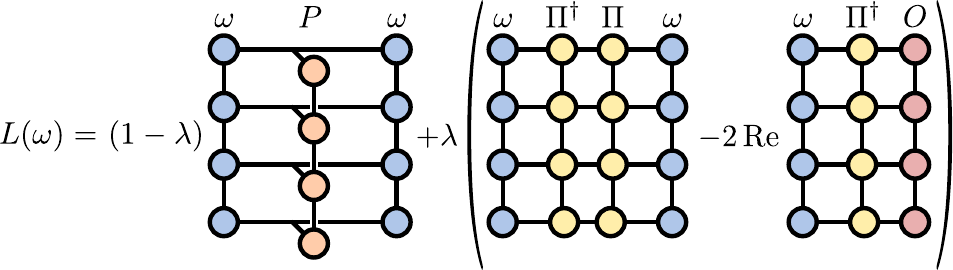}
\end{equation*}
The \emph{global} cost function $L(\omega)$ is a quadratic function of the estimator tensor $\omega$ and, as we will now show, its minimization can be reduced to a sequence of \emph{local} quadratic problems whose solutions are obtained by solving linear systems of equations. Of course, reducing the minimization of a global function to a sequence of local ones comes at the risk of encountering local, rather than global, minima of the cost. Nonetheless, such iterative methods common in the tensor network literature~\cite{Schollwok2011DMRGMPS, GuoQEMviaMPO2022}, are in practice found to converge to good solutions.

Let $\bm{\omega}_\ell$ denote the local tensor at site $\ell$ in the MPS $\kket{\omega}$. Suppose we fix all the tensors $\bm{\omega}_j$ at the remaining sites $j \neq \ell$, then the cost function $L(\omega)$ in terms of only the $\ell-$th local tensor amounts to
\begin{equation}
\begin{aligned}
    \label{eq:local-cost}
    L\qty(\bm{\omega}_\ell) = (1-&\lambda)\, \bm{\omega}_\ell^\intercal A_\ell \bm{\omega}_\ell + \\ 
    + & \lambda\, \qty(\bm{\omega}_\ell^\intercal B_\ell \bm{\omega}_\ell - 2\text{Re}\qty[\bm{\omega}_\ell^\intercal \bm{v}_\ell])\,,
\end{aligned}
\end{equation}
where $A_\ell$, $B_\ell$ and $\bm{v}_\ell$ are the so-called environment tensors obtained by contracting all the tensors but $\bm{\omega}_\ell$ in the tensor networks in Eq.~\eqref{eq:tn-cost-fun}, and $(\bm{\cdot})^\intercal$ denotes transposition. We use a bold notation for $\bm{\omega}_\ell$ and $\bm{v}_\ell$ to indicate that these tensors behave like vectors, while $A_\ell$ and $B_\ell$ instead act like matrices. These can be obtained by an appropriate reshaping of the corresponding tensors, as clear from the diagrammatic representation
\begin{equation*}
    \includegraphics[width=\columnwidth, valign=c]{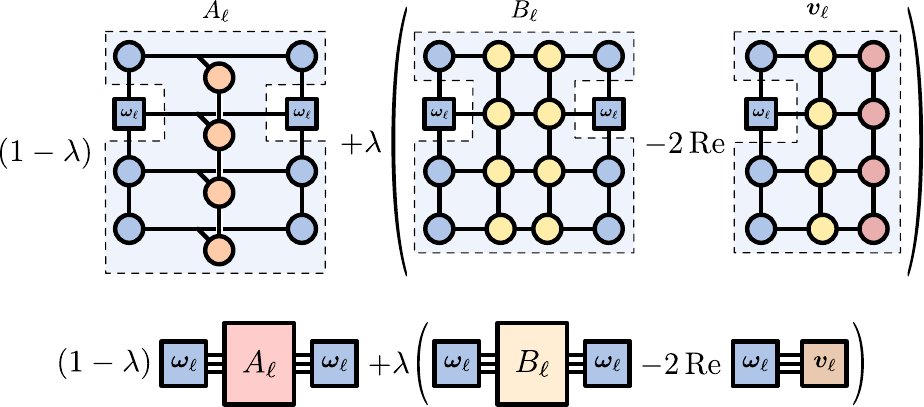}
\end{equation*}
The matrices $A_\ell$ and $B_\ell$ are of size $(s\chi^2, s\chi^2)$, where $\chi$ is the bond dimension of the MPS estimator, and $s$ is the dimension of the local sites in the estimator, defined by the number of outcomes per qubit associated to the measurement process (for example, $s=6$ for an OC-POVM consisting of 6 possible outcomes per qubit). 
Also, we can parameterize the MPS estimator to contain only real entries, so that the linear term in Eq.~\eqref{eq:local-cost} can be further simplified to $\Re\qty[\bm{\omega}_\ell^\intercal \bm{v}_\ell] = \bm{\omega}_\ell^\intercal \Re\qty[\bm{v}_\ell]$.

One can readily realize that the local cost function $L(\bm{\omega}_\ell)$ in~\eqref{eq:local-cost} is again a quadratic form with respect to the local variables $\bm{\omega}_\ell$, and its minimum is found by solving the linear system of equations~\cite{MatrixCookbook}
\begin{equation}
\begin{aligned}
    \label{eq:local-cost-solution}
    & \bm{\omega}_\ell^{\text{opt}} = \argmin_{\bm{\omega}_\ell} L(\bm{\omega}_\ell) \\
    & \qty[(1-\lambda)(A_\ell + A_\ell^\intercal) + \lambda (B_\ell + B_\ell^\intercal)]\bm{\omega}_\ell^{\text{opt}} = 2 \lambda \Re \bm{v}_\ell
\end{aligned}\!.
\end{equation}

As customary in tensor network procedures, we variationally search for the minimum of the global cost function $L(\omega)$~\eqref{eq:tn-cost-fun} by sweeping back and forth over the sites $\ell$ of the MPS and solving the local quadratic problems $L(\bm{\omega}_\ell)$~\eqref{eq:local-cost} using the explicit solution Eq.~\eqref{eq:local-cost-solution}. 

Summarizing, we have shown how to express the problem of finding a low-variance observable estimator as an optimization task defined in terms of tensor networks. Also, we have proposed an efficient solution to such an optimization problem by reducing it to the task of sequentially solving linear systems of equations for the local sites in the tensor network. 

We refer to Appendix~\ref{app:optimization_details} for numerical details about the method.

\section{Performance and reconstruction guarantees} 
\label{sec:unbiased-estimation}
If the penalty-regularized variance minimization process is successful, at the end of the optimization we obtain a tensor estimator attaining low statistical variance $\Var{\omega}$, and small reconstruction error $\norm{O-O_\omega}_2 = \varepsilon \ll 1$, with $O_{\omega} = \sum_k \omega_k \Pi_k$ being the approximate reconstruction of the target observable $O$. The approximate reconstruction induces an estimation bias $\abs{\Tr[O_\omega\rho]- \Tr[O\rho]}\neq0$ in the estimation. In the case of infinite statistics, we can show that (see Appendix~\ref{app:performance-guarantee})
\begin{equation}
\begin{aligned}
    \label{eq:bias-bound}
    \abs{\expval{O_\omega} -\expval{O}} &= \abs{\Tr[O_\omega\rho]- \Tr[O\rho]} \leq \varepsilon\,, \quad \forall \rho\,.
\end{aligned}
\end{equation}

It is also possible to obtain performance guarantees of the optimized estimator in the finite statistics case. Let $\overbar{\omega}$ denote the empirical mean estimator obtained with $S$ measurement shots with the optimized tensor estimator
\begin{equation}
    \label{eq:mps-empirical-average}
    \overbar{\omega} \coloneqq \frac{1}{S}\sum_{s=1}^S \omega_{k_s}\,,
\end{equation}
where $\omega_{k_s}$ denotes the reconstruction coefficient labeled $k_s$ observed in the $s$-th experimental measurement shot. We show in Appendix~\ref{app:performance-guarantee} that the probability that the empirical average $\overbar{\omega}$ is far to the true observable expectation value $\expval{O}$ can be upper bounded with a Chebyshev-like inequality as
\begin{equation}
    \label{eq:performance-guarantee}
    \begin{aligned}
        \text{Pr}\qty(\abs{\overbar{\omega} - \expval{O}} > \delta) \leq \frac{\Var{\omega}}{\delta^2 S} + \frac{\varepsilon^2}{\delta^2}\,,
    \end{aligned}
\end{equation}
The bound is only informative as long as the reconstruction error is less than the required accuracy, namely $\delta > \varepsilon$.

As one would expect, the bound in~\eqref{eq:performance-guarantee} consists of two qualitatively different terms: the first term relates to the statistical fluctuations in the estimation procedure, depending on the estimator variance $\Var{\omega}$ and the number of measurement shots $S$ in the sample mean $\overbar{\omega}$. The second term, on the other hand, does not depend on the statistical uncertainty but takes into account the fact that the tensor estimator provides only an $\varepsilon$-close approximation of the true observable. 

Notably, tighter concentration bounds with Hoeffding-like performances can also be derived under specific assumptions (\ie sub-gaussian estimator) valid for specific observables and measurement strategies (\ie the POVM effects), using the same techniques proposed in~\cite{ShadowTomographyDualInnocenti2023, ThriftyShadowEstimation2023, FermionicShadowTomography2021, AcharyaShadowTomographyPOVM2021}. In particular, we show in Appendix~\ref{app:performance-guarantee} that such improved concentration bounds can be derived also in the case of biased estimation, both for the sample mean~\eqref{eq:mps-empirical-average} and for the median-of-means estimator originally proposed in~\cite{HuangShadows2020}. We refer to Appendix~\ref{app:performance-guarantee} for an in-depth discussion on the topic, and here restrict our attention to using the sample mean estimator~\eqref{eq:mps-empirical-average}, and showing one prototypical example of how concentration bounds can be straightforwardly derived also for biased estimators, as reported in Eq.~\eqref{eq:performance-guarantee}.

\section{Numerical results}
\label{sec:numerics}
In this section, we report numerical results for several observable estimation tasks consisting of different system sizes. In the numerical examples, we will compare TN-ICE with classical shadows/canonical dual-frame estimator. In the first subsections, we consider the infinite statistics scenario, while in the last subsection we study the effect of finite statistics. 

In all numerical experiments, we consider the common randomized measurement strategy consisting of measuring the Pauli observables $X$, $Y$ and $Z$ with equal probability on each qubit. Such measurement protocol corresponds to associating each qubit with an informationally over-complete (OC) POVM whose effects are
\begin{equation}
\label{eq:oc-pauli-effects}
    \begin{aligned}
    \pi_k \in \Biggl\{ & \frac{1}{3} \dyad{0},~ \frac{1}{3} \dyad{1},~ \frac{1}{3} \dyad{+}, \\
    & \frac{1}{3} \dyad{-},~ \frac{1}{3} \dyad{+i},~ \frac{1}{3} \dyad{-i} \Biggr\}\,.
    \end{aligned}
\end{equation}
The multi-qubit OC-POVM is then given by tensor products of the local effects $\{\Pi_k\}_k = \{\pi_{k_1} \otimes \ldots \otimes \pi_{k_n}\}_{k_1,\ldots,k_n}$, thus consisting of a total of $r = 6^n$ effects.

Unless otherwise specified, before the optimization process~\eqref{eq:local-cost-solution} begins the MPS estimators $\omega$ are initialized with a random MPS with normally distributed entries. The data for reproducing the chemical examples discussed in Sec.~\ref{sec:chem-examples} can be found at~\cite{tnice-data-chem-zenodo}.

\subsection{GHZ states}
\label{sec:res-ghz-infinite}
We start with the example of estimating the observable $O = X^{\otimes n} - Y^{\otimes n}$ using the GHZ state
\begin{equation}
    \label{eq:ghz_state}
    \ket{\text{GHZ}} = \ket{\psi} = \frac{\ket{0}^{\otimes n} + \ket{1}^{\otimes n}}{\sqrt{2}}\,.
\end{equation}
For $n$ even the state is an eigenstate of the observable so that the observable variance $\smash{\Var{O} \coloneqq \expval{O^2} - \expval{O}^2}$ vanishes. 

\begin{figure}[!ht]
    \centering
    \includegraphics[width = \columnwidth]{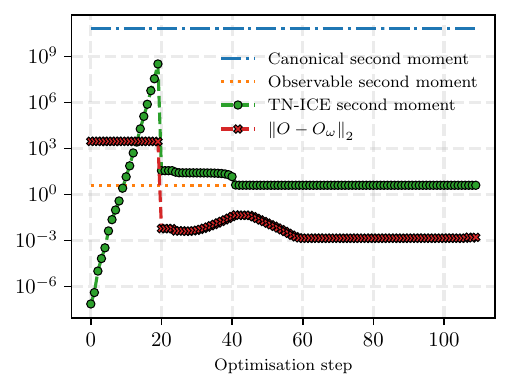}
    \caption{Iterative minimization of the cost~\eqref{eq:local-cost} for a system of $n=22$ qubits, with MPS bond dimension $\chi = 8$ and regularization coefficient $\lambda = 0.999$. The measured state is an eigenstate of the observable to be estimated with eigenvalue $2$ so that $\expval{O}=2$ and $\expval{O^2} = 4$. Observable second moment refers to $\expval{O^2}$, canonical to the second moment of the canonical estimator with canonical duals (classical shadows), while TN-ICE is the second moment of the proposed estimator obtained at the end of the penalty-regularized variance minimization procedure. After optimization TN-ICE reaches a low second moment and reconstruction error $\big\|O - O_{\omega}\big\|_2$, thus providing an accurate and unbiased estimation.}
    \label{fig:ghz_example}
\end{figure}

In Fig.~\ref{fig:ghz_example} we report results for a system composed of $n=22$ qubits, with a maximum bond dimension of the MPS estimator of $\chi = 8$, and with the regularization coefficient in the cost function~\eqref{eq:local-cost} set to $\lambda = 0.999$. 
In the plot we also report the estimator second moment corresponding to classical shadows/canonical duals (see Eq.~\eqref{eq:app-classical-shadows} in Appendix~\ref{app:dual-frames-classical-shadows} for more details), which for global observables scales exponentially with the system size~\cite{Malmi2024EnhancedEstimation, HuangShadows2020}. Additionally, we also report the lowest second moment possible which would be achieved by measuring the state in the eigenbasis of the observable. This amounts to $\E{O^2} = \Tr[O^2 \dyad{\psi}] = 4$, since the state is an eigenstate with eigenvalue 2.

Using the sweeping iterative local optimization routine~\eqref{eq:local-cost}, we can see that both the second moment $\E{\omega^2}$ and the penalty term $\norm{\kket{O} - \Pi \kket{\omega}}_2$ are minimized during training. After only a couple of sweeps through the MPS chain, we can find a set of coefficients $\{\omega_k\}_k$ which not only satisfy the observable reconstruction constraint with low error, hence providing an unbiased estimator, but most importantly match the performance of the best possible estimation strategy.  

This example shows that, even when performing a local measurement, our correlated tensor network estimation procedure can capture the correlations in the measurement data and post-processing them to provide a near-zero variance estimation, thereby identifying that the measured state is an eigenstate of the observable.

\subsection{Chemical examples}
\label{sec:chem-examples}
We now showcase the efficacy of TN-ICE in estimating the energy of the ground states of chemical Hamiltonians. Specifically, we consider the molecules LiH, N$_{2}$ and H$_{6}$ mapped on qubit systems of size $n=12$, and whose ground states are prepared with a quantum circuit using the ADAPT-VQE~\cite{GrimsleyADAPT2019, qAdaptTang2021} technique, with convergence tolerance set to $10^{-3}$ Hartree close to exact diagonalization of the Hamiltonian. We refer to ref.~\cite{Miller2024treespilation} for more details on the ground state preparation of such molecules.

In Table~\ref{tab:chemical-summary} we report the energy estimator variance of the tensor network estimator after optimization, compared with the variance obtained with the estimator built with fixed tensor product canonical duals/classical shadows, and with the observable variance $\Var{O}$, which represents a lower bound to any estimation procedure. Note that the observable variance is small but non-zero because the quantum circuits are only approximate representations of the ground states of the Hamiltonians. 
\begin{table}[ht]
\setlength{\tabcolsep}{2.5pt}
\begin{tabular}{l|cccc}
& & LiH & N$_{2}$ &  H$_{6}$ \\[0.1em] \hline \\[-0.8em]
\multirow{3}{*}{Variance}   & Observable &$ 6\!\times\! 10^{-4}$ & $2.2\!\times\! 10^{-3}$ & $1.2\!\times\! 10^{-3}$ \\
                            & Canonical & 298.98 & 467.40 & 1301.75 \\
                            & \textbf{TN-ICE} & \textbf{0.77} & \textbf{9.35} & \textbf{62.91} \\[0.1em] \hline \\[-0.8em]
Penalty & & $4\!\times\!10^{-4}$ & $3\!\times\!10^{-4}$ & $9\!\times\!10^{-4}$
\end{tabular}
\caption{Performances of using TN-ICE for minimizing the estimator variance for the energy of three molecules. Observable refers to the observable variance $\Var{O}$, Canonical to the use of the canonical estimator with canonical duals (classical shadows), while TN-ICE is our estimator obtained at the end of the penalty-regularized variance minimization procedure. The penalty is the estimator penalty $\norm{\kket{O}-\Pi\kket{\omega}}_2$ introduced in the cost~\eqref{eq:cost-function}. The number of qubits used to represent the ground states of the molecules is $n=12$, the bond dimension used in the optimisation is $\chi=60$, and $\lambda=0.9999$, see Eq.~\eqref{eq:cost-function}.}
\label{tab:chemical-summary}
\end{table}

In all considered cases TN-ICE is able to drastically reduce the estimator variance by at least two orders of magnitude, importantly with a small reconstruction error and reconstruction error $\big\|O - O_{\omega}\big\|_2$ that ensures a faithful reconstruction of the observable. The MPS $\kket{\omega}$ used in these simulations has bond dimension $\chi=60$.

\begin{figure}[ht]
    \centering
    \includegraphics{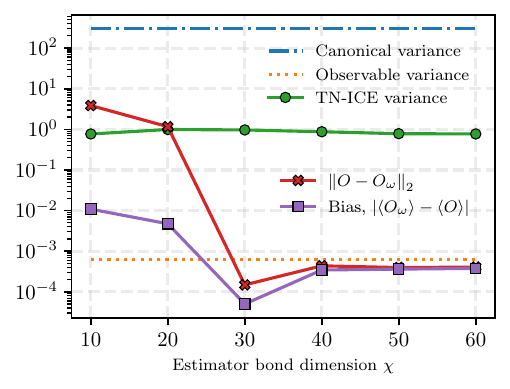}
    \caption{Effect of the bond dimension $\chi$ of the statistical performances of the estimator at the end of training for estimating the energy of LiH on a system of $n=12$. While for small bond dimension the estimator is unable to represent the observable in the effects basis (\ie we observed a large and reconstruction error $\big\|O - O_{\omega}\big\|_2$), for large enough bond dimension we can obtain a faithful estimation with small variance.}
    \label{fig:lih-bonddim-scaling}
\end{figure}

In Fig.~\ref{fig:lih-bonddim-scaling} we study the effect of the estimator bond dimension $\chi$ on the performances for the molecule LiH. Specifically, in the plot we report the estimator variance, the penalty term and the estimation bias $\abs{\expval{O_\omega} - \expval{O}}$ obtained at the end of training for increasing bond dimension. Note that, as discussed in Sec.~\ref{sec:unbiased-estimation}, the bias is always upper bounded by the reconstruction error, \ie $\abs{\expval{O_\omega}-\expval{O}}\leq \big\|O - O_{\omega}\big\|_2$.

For small values of the bond dimension, $\chi \leq 20$, the penalty term is large, thus signalling that the estimator $\omega$ is unable to capture the correlations necessary to express the observable in the effects basis (see Eq.~\eqref{eq:operator-decomposition}). This implies that such estimators cannot provide reliable estimations in general, despite its variance and error being already small in this particular case.

However, provided enough bond dimension $\chi \geq 30$, it is possible to obtain a faithful reconstruction of the observable (low reconstruction error) achieving much smaller variance compared to classical shadows. In particular, note that the transition point happens around $\chi\approx30$ which, in this case, is the bond dimension necessary to represent the  Hamiltonian in MPS form, hence sufficient to build the canonical estimator coefficients $o_k = \Tr[O D_k] = \bbrakket{O}{D_k}$ using tensor product canonical duals $D_k$. Compared with classical shadows using fixed uncorrelated duals, this shows that with comparable resources our optimized tensor network estimator can leverage the redundant degrees of freedom in the overcomplete measurements to reduce the statistical fluctuations of the estimation process.

\subsection{Finite statistics}
\label{sec:finite-ghz}
All previous examples were performed using the quantum probabilities $p_k = \Tr[\Pi_k \rho]$, which can only be obtained using an infinite measurement budget. We now show how the proposed method applies also when one has access only to a limited number $S$ of measurement outcomes, and hence deals with observed frequencies $f_k$ rather than the probabilities $p_k$ when computing the cost functions~\eqref{eq:tn-cost-fun}. 

\begin{figure}[ht]
    \centering
    \includegraphics[width=\columnwidth]{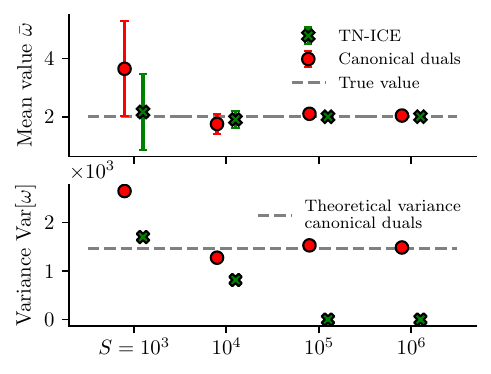}
    \caption{Using TN-ICE in the case of finite statistics. We report the results obtained at the end of optimization with a training dataset of varying sizes $S = \{10^3, 10^4, 10^5, 10^6\}$, and then checking the performances on an independent test set of the same dimension. In the panel above it is shown the empirical mean estimator $\overbar{\omega}$~\eqref{eq:mean-estimator} for the optimized tensor estimator and the canonical duals, with the error bars indicating one standard error $\sqrt{\text{Var}[\omega]/S}$~\eqref{eq:std}. In the panel below, we report the corresponding estimator variances $\text{Var}[\omega]$~\eqref{eq:estimator-variance}. As more data is available, both methods converge to the true expectation value, but the optimized tensor estimator is more accurate and eventually achieves a zero-variance estimation. The full optimization runs for all the data points in these plots are reported in Fig.~\ref{fig:full-sim-ghz} in Appendix~\ref{app:additional-results}.}
    \label{fig:finite_stats_ghz}
\end{figure}

We investigate finite statistics on the observable estimation task on a GHZ state already introduced in Sec.~\ref{sec:res-ghz-infinite}, but on a system of $n = 6$ qubits, and study its performances with a varying number of measurement shots $S \in \{10^3, 10^4, 10^5, 10^6\}$. In Fig.~\ref{fig:finite_stats_ghz} we report the results obtained with the proposed optimized tensor network estimator with bond dimension $\chi = 8$ and those obtained with canonical duals (classical shadows) on the same datasets.

From the picture it is clear that while both methods converge to the true expectation value when enough samples are available, the tensor network is able to provide an estimation with less statistical error, eventually achieving a zero-variance estimation for large sample sizes, similar to the results shown previously in the case of infinite statistics, see Sec.~\ref{sec:res-ghz-infinite}. 

Note that this specific case is an example of a particularly hard estimation task for randomized measurement schemes, since it involves predicting the expectation value of a global observable $O = X^{\otimes n} - Y^{\otimes n}$ using only local single-qubit randomized measurements~\eqref{eq:oc-pauli-effects}. This scenario is known to scale badly since the canonical estimator coefficients can lie in an exponentially large interval $\omega_k^{\text{can}} \in [-a^n, a^n],\, a>1$, which translates in an exponentially large estimation variance and hence an exponential amount of measurement to reach good accuracy~\cite{HuangShadows2020, Malmi2024EnhancedEstimation}. Such lack of information imposed by a finite measurement budget cannot be solved by classical post-processing alone, so we expect our method to also require an exponentially large amount of data whenever local measurements are used to estimate global observables. In practice, this means that while we can reach optimal estimation performances with $S=10^6$ shots on $n=6$ qubits in Fig.~\ref{fig:finite_stats_ghz}, more shots will be needed for larger systems. Such bottleneck can be addressed for example by using global, instead of local, random measurement strategies~\cite{Hu2021ShallowShadow, HuangShadows2020}, which can be naturally implemented in our tensor estimation framework provided that the correlated POVM has an efficient tensor network description.

One important point concerning optimisation problems with finite statistics is to guarantee that the process does not lead to overfitting. This occurs when the dataset used for the optimisation is small and it is not representative of the underlying distribution. A typical strategy to detect this phenomenon is to use a cross-validation procedure with two statistically independent datasets, one that is used for optimisation (called training set) and another one to provide the final estimation (called test set). During the optimisation process, we compute the value of the optimised estimator for both datasets. Whenever the performance of the estimator on the test set starts deteriorating, then the minimization procedure is stopped. In Fig.~\ref{fig:full-sim-ghz} in Appendix~\ref{app:additional-results} we report the optimization runs for both training and test datasets used to generate the data in Fig.~\ref{fig:finite_stats_ghz}. As one can see, while the estimation on the training dataset always decreases (because of the minimisation procedure), it can increase on the test dataset, indicating overfitting.

Another point is that while for large dataset sizes the optimization process is always capable of converging to a good estimator with low variance and negligible reconstruction error, regardless of the initial estimator at the start of training, this is not the case for finite statistics. Indeed, in order to have a meaningful optimization process, whenever the amount of data is scarce we observe that is desirable to start the training from an estimator having a correct first moment and a small reconstruction error, obtained for example initializing it to the canonical estimator. Indeed, for the results reported in Fig.~\eqref{fig:finite_stats_ghz} and in Fig.~\ref{fig:full-sim-ghz} in Appendix~\ref{app:additional-results}, the estimator was initialized to the canonical one for $S=10^3, 10^4$, to a perturbed version of the canonical one obtained by adding random normal noise to its entries for $S=10^5$, and a random MPS for $S=10^6$.

Overall, our simulations then suggest that the tensor-network observable estimation is capable of providing unbiased and low-variance estimations even in the finite statistics regime. 

\section{Conclusions}
\label{sec:final-remarks}
In this manuscript we have introduced TN-ICE, a method for low-variance estimations of the expectation values of many observables that uses tensor networks to post-process data from informationally (over-)complete measurements. For each observable of interest, the technique works by finding a decomposition of the target observable in terms of the POVM effects that results in the lowest estimation variance. This is opposed to standard classical shadows techniques where the post-processing method is fixed and only depends on the measurement strategy.

We have shown how to do this efficiently by first parameterizing the observable's reconstruction coefficients as an MPS, and then optimizing it with a DMRG-like minimization procedure that involves solving local systems of linear equations. Our method can be used to find reliable low-variance estimators with provable guarantees for arbitrary measurement strategies ---provided they define an informationally complete measurement--- and observables that can be efficiently represented in terms of tensor networks.

By classically post-processing the measurement outcomes with tensor networks, our estimator can capture global correlations in the experimental data, even when using measurements that only act locally on each qubit. This enables low-variance estimations which ultimately can help reducing the measurement overhead coming with observable estimation tasks. We demonstrated the efficacy of the proposed method with several numerical experiments, as reported in Sec.~\ref{sec:numerics}.

We also emphasize that TN-ICE can converge to the optimal estimator in the limit of a large bond dimension. This follows from the fact that, by sufficiently increasing the bond dimension of the MPS estimator, we can accurately approximate the state-dependent optimal dual frame prescribed in~\cite{ShadowTomographyDualInnocenti2023, OptimalProcessingDariano2007}. Indeed, in some of our examples, we have observed that TN-ICE approaches the optimal state-dependent variance.

We now highlight some interesting directions for future research on the topic. Although TN-ICE is generally applicable to any informationally complete (IC) measurement, in this work we have tested it only on single-qubit Pauli measurements. As future direction, it would be interesting to test it with more complex measurement strategies, including recent proposals based on the shallow quantum circuits~\cite{Hu2021ShallowShadow, Arienzo2022ShallowShadow, Bertoni2023shallow, Akhtar2023ShallowShadowTN, Hu2024RobustShallowShadow, Farias2024RobustShallowShadow}. This approach opens up a whole new venue for investigating effective measurement strategies that could further reduce the statistical error of the estimation. 

Additionally, while we have numerically confirmed that TN-ICE can work also with finite statistics data on a small scale example, further studies are needed to asses the impact of overfitting and explore potential methods for mitigating it. Regarding the optimization procedure of the MPS, it could be valuable to investigate some of the techniques proposed in the tensor network literature for similar optimization problems~\cite{DolgovDMRGTTN2013}. As an example, using two-sites optimization routines could help mitigate the insurgence of local minima and allow for a variable bond dimension during optimization.

Finally, we believe that our method could be combined with other strategies to minimize errors in quantum computations. For instance, it could be integrated with error mitigation methods, such as the tensor-network error mitigation approach recently presented in Ref.~\cite{TEMFilippov2023}.

\section{Acknowledgement}
S.M and D.C thank Guillermo García-Pérez, Joonas Malmi, Stefan Knecht, Keijo Korhonen and Luca Innocenti for helpful discussions. S.M. thanks Dario Gasbarra for useful discussions on concentration bounds. Work on “Quantum Computing for Photon-Drug Interactions in Cancer Prevention and Treatment” is supported by Wellcome Leap as part of the Q4Bio Program.

\section{Competing interests}
Elements of this work are included in a patent filed by Algorithm Ltd with the European Patent Office.

\bibliographystyle{quantum}
\bibliography{main}

\appendix
\onecolumngrid

\section{Vectorized or double-ket notation}
\label{app:vectorisation}
In this section, we briefly introduce the vectorized, or double-ket, notation for linear operators. Given a linear operator $O$, one can define the vectorized operator $\kket{O}$ defined as~\cite{OptimalProcessingDariano2007, Wood2015tensornetworksgraphicalcalculus}
\begin{equation}
    O = \sum_{i, j} O_{ij} \ketbra{i}{j} \longrightarrow \kket{O} = \sum_{i,j} O_{ij} \ket{i}\otimes\ket{j}\,.
\end{equation}
where the operator and its vectorized vector have been expressed in the computational basis. Another common way of vectorizing an operator is to use the Pauli basis since this forms a basis in the space of complex matrices. In this case one has 
\begin{equation}
    \label{eq:app-ptm-vec}
    O \longrightarrow \kket{O} = \frac{1}{\sqrt{2^n}}\sum_{k=1}^{4^n}\Tr[O^\dagger P_k] \kket{P_k}
\end{equation}
where $P_k = P_{k_1} \otimes \ldots P_{k_n}$, with $P_{k_i} \in \{\mathbb{I}, X, Y, Z\}$ being the single-qubit Pauli matrices. Such representation is usually referred to as Pauli Transfer Matrix (PTM) representation, especially when used to represent quantum channels as matrices~\cite{Greenbaum2015introductionquantumgateset, ManginiDeconvolution2022, Roncallo2023PTM}.

\section{Equivalence of the classical shadows and dual frames for local Pauli measurements}
\label{app:dual-frames-classical-shadows}
In this Appendix we work out explicitly the correspondence between the recently introduced classical shadows~\cite{HuangShadows2020, Paini2021estimating} and the formalism of quantum measurement frames, which have been used in the literature already for some time already~\cite{ShadowTomographyDualInnocenti2023, ScottTightICPOVM2006, OptimalProcessingDariano2007, DArianoICMeasurements2004, ZhouOCPOVM2014}. Such equivalence has been already discussed in depth in~\cite{ShadowTomographyDualInnocenti2023, AcharyaShadowTomographyPOVM2021} and so, for the sake of simplicity, we hereby work out explicitly such correspondence only for a specific practical example, namely the common and easily implementable protocol of randomized single-qubit Pauli measurements. In particular, fixed the measurement primitive (\ie a POVM) for the shadow protocol, we will show how the so-called classical shadows are nothing more than a specific choice of dual frame to the chosen POVM, dubbed \textit{canonical duals} in the frame literature.

\subsection{Estimation in the formalism of classical shadows}
Let's start by summarizing the steps required to run the classical shadow protocol~\cite{HuangShadows2020}, where without loss of generality we use the sample mean estimator instead of the originally proposed median-of-means (see discussions in Sec.~\ref{app:sec_hoeffding} on why the sample mean is sufficient).
\begin{algorithm}
\caption{Description of the classical shadow estimation protocol~\cite{HuangShadows2020}.}
\KwIn{Quantum state $\rho$;}
\myinput{Observable $O$;}
\myinput{Measurement primitive defined by a collection of unitaries $\mathcal{U} = \{U_i\}_i$;} 
\myinput{Number of shots $S$.}
\KwResult{Estimate $\overbar{o}$ of $\expval{O} = \Tr[O\rho]$.}
    \For{$s = 1$ to $S$}{
    Draw a random unitary $U_s$ from $\mathcal{U}$\;
    Apply unitary to state $\rho \rightarrow U_s \rho U_s^\dagger$ and measure in the computational basis, store result $\hat{b}_s \in \{0,1\}^n$\;
    Define the measurement channel operator 
    \begin{align}
        \mathcal{M}(\rho) \coloneqq \mathbb{E}_{U, b}\qty[U^\dagger\!\dyad{b}U] = \mathbb{E}_{U}\mathbb{E}_b\qty[U^\dagger\!\dyad{b}U] & = \sum_{U \in\, \mathcal{U}} p(U) \sum_{b \in \{0,1\}^n} p(b)~U^\dagger\!\dyad{b}U \\
        & = \sum_{U \in\, \mathcal{U}} \sum_{b \in \{0,1\}^n} p(U) \mel{b}{U\!\rho U^\dagger}{b}\,U^\dagger\!\dyad{b}U
    \end{align}
    where $p(U)$ is the probability of sampling unitary $U$, and $p(b) = \mel{b}{U\!\rho U^\dagger}{b}$ is the probability of measuring bitstring $b$ on state $U\!\rho U^\dagger$\; 
    Compute and store the \textit{classical shadow} of the state $\hat{\rho}_s \coloneqq \mathcal{M}^{-1}(U_s^\dagger\dyad{b_s}U_s)$\;
    }
    Given the collection of classical shadows $\{\hat{\rho}_1, \ldots, \hat{\rho}_S\}$, compute the estimators $o_s = \Tr[\hat{\rho}_s O]$\;
    \KwResult{Compute the sample mean $\overbar{o} = \frac{1}{S}\sum_{i=1}^S o_s$}
\end{algorithm}

\noindent Note that for simplicity we assumed that the collection of unitary is discrete $\mathcal{U} = \{U_i\}_i$, each being sampled with probability $p(U_i)$. In the case of a continuous set of unitaries, the sums are substituted with appropriate integrals. Also, note that in the case of median-of-means estimation, one only modifies the last step in the protocol by clustering the snapshots $\hat{\rho}_s$ in disjoint subsets, and then computing the median of the means obtained in each subset.

Consider the common and readily implementable measurement primitive consisting of randomized single-qubit Pauli measurements, defined by randomly measuring each qubit on the $X$, $Y$ or $Z$ basis with equal probability. Since the measurements are local, we focus on a single-qubit case but one can then easily generalize to multi-qubit systems simply by constructing tensor products of the single-qubit shadows. 

The single-qubit randomized Pauli measurement primitive is implemented by applying on each qubit $\rho \rightarrow U \rho U^\dagger$ a random unitary $U$ from the collection 
\begin{equation}
    \label{eq:app_random_pauli_shadow}
    U \in \mathcal{U} = \{\mathbb{I}, H, HS^\dagger\}
\end{equation}
that allows for changing the measurement basis from the eigenstates of Pauli-$Z$ to that of $X$ and $Y$ since,
\begin{equation}
    HZH = X \quad SHZHS^\dagger = Y\,.
\end{equation}
Importantly, note that such measurement protocol is equivalent to measuring the qubit according to the POVM with effects
\begin{equation}
\label{eq:app-oc-pauli-effects}
    \Pi_0 = \frac{1}{3} \dyad{0},~ \Pi_1 = \frac{1}{3} \dyad{1},~ \Pi_2 = \frac{1}{3} \dyad{+},~ \Pi_3 = \frac{1}{3} \dyad{-},~ \Pi_4 = \frac{1}{3} \dyad{+i},~ \Pi_5 = \frac{1}{3} \dyad{-i}\,.
\end{equation}
consisting of the six (sub-normalized) Pauli eigenstates. One can easily check that $\Pi = \{\Pi_k\}_k$ is indeed a valid POVM since $\sum_k \Pi_k = \mathbb{I}$ and $\Pi_k \geq 0$. In particular, such POVM is not only \textit{informationally complete} (IC) ---that is, its effects form a basis in operator space---, but most importantly it is \textit{overcomplete} (OC), since the number of effects (6) is larger than the single qubit operator space dimension (4), so any operator decomposition in this basis will feature some redundant degrees of freedom.

Let's proceed with computing explicitly the measurement channel $\mathcal{M}$ induced by the choice of the unitaries~\eqref{eq:app_random_pauli_shadow}. First of all, let's rewrite the channel in a more convenient form,
\begin{align}
    \mathcal{M}(\rho) &= \sum_{U \in\, \mathcal{U}} \sum_{b} p(U) \mel{b}{U\!\rho U^\dagger}{b}\,U^\dagger\!\dyad{b}U \nonumber \\
    &= \frac{1}{3}\sum_{U \in\, [H, HS^\dagger, \mathbb{Id}]} \sum_{b \in \{0,1\}^n} \Tr[\rho~ U^\dagger\!\dyad{b}U]\,U^\dagger\!\dyad{b}U\,, \label{eq:app_explicit_meas_channel}
\end{align}
where we have used $p(U) = 1/3~\forall U \in \mathcal{U}$ since all unitaries are equally probable, and expressed the bitstring probability in the form of a trace, so that the expression now only depends on the operators $U^\dagger\dyad{b}U$. Then, using
\begin{equation}
\label{eq:app-possible_combinations_shadow}
\begin{aligned}
    \dyad{0} = \frac{\mathbb{I} + Z}{2}\,, 
    \quad\quad
    H \dyad{0} H & = \dyad{+} = \frac{\mathbb{I} + X}{2}\,, \quad\quad 
    SH \dyad{0} HS^\dagger = \dyad{+i} = \frac{\mathbb{I} + Y}{2}\,, \\
    \dyad{1} = \frac{\mathbb{I} - Z}{2}\,, 
    \quad\quad
    H \dyad{1} H & = \dyad{-} = \frac{\mathbb{I} - X}{2}\,, 
    \quad\quad  
    SH \dyad{1} HS^\dagger = \dyad{-i} = \frac{\mathbb{I} - Y}{2}\,,
\end{aligned}
\end{equation}
and that fact that the Pauli matrices $\{\mathbb{I}, X, Y, Z\}$ form a basis in the space $2 \times 2$ complex matrices, 
\begin{equation}
    A = \frac{\Tr[A]\mathbb{Id} + \Tr[A X]X + \Tr[A Y]Y + \Tr[A Z] Z}{2}\,,
\end{equation}
in Eq.~\eqref{eq:app_explicit_meas_channel}, one can perform explicitly the summations and finally obtain
\begin{equation}
    \mathcal{M}(\rho) = \frac{1}{3}\qty(\rho + \Tr[\rho]\mathbb{I}) = \frac{1}{3}\rho + \frac{2}{3}\Tr[\rho]\frac{\mathbb{I}}{2}.
\end{equation}
We thus realize that the measurement channel is a depolarizing channel $\mathcal{D}_p(\rho) = p \rho + (1-p)\Tr[\rho]\mathbb{I}/2$ with intensity $p = 1/3$. Such channels can be readily inverted as~\cite{ManginiDeconvolution2022}
\begin{equation}
    \mathcal{D}_p^{-1}(\rho) = \frac{1}{p}A - \frac{1-p}{p}\Tr[A]\frac{\mathbb{I}}{2}\,,
\end{equation}
and specifically for $p=1/3$ one has $\mathcal{D}_{1/3}^{-1}(\rho) = 3\rho - \Tr[\rho]\mathbb{I}$. 

Using this inversion formula one can then compute the single-shot classical shadows $\hat{\rho}_s = \mathcal{M}^{-1}(U_S^\dagger\dyad{b_s}U_s)$. In particular, notice that the measurement primitive~\eqref{eq:app_random_pauli_shadow} implies that there are only 6 possible values for the classical snapshot $\rho_s$, obtained as the possible combinations of three change of basis unitaries $U_s \in \{\mathbb{I}, H, HS^\dagger\}$ with the two possible measurement outcomes $b_s = \{0,1\}$, see for example Eqs.~\eqref{eq:app-possible_combinations_shadow}. Applying the inverse channel $\mathcal{M}^{-1} = \smash{\mathcal{D}_{1/3}^{-1}}$ on the six possible operators $U_s^\dagger\dyad{b_s}U_s$ in Eqs~\eqref{eq:app-possible_combinations_shadow}, one then has
\begin{equation}
\begin{aligned}
    \label{eq:app-classical-shadows}
    \hat{\rho}_0 &= \mathcal{M}^{-1}\qty(\frac{\mathbb{I}+Z}{2}) = \frac{\mathbb{I} + 3Z}{2}\,,
    \quad\quad
    \hat{\rho}_2 = \mathcal{M}^{-1}\qty(\frac{\mathbb{I}+X}{2}) = \frac{\mathbb{I} + 3X}{2}\,,
    \quad\quad
    \hat{\rho}_4 = \mathcal{M}^{-1}\qty(\frac{\mathbb{I}+Y}{2}) = \frac{\mathbb{I} - 3Y}{2}\,, \\
    \hat{\rho}_1 &= \mathcal{M}^{-1}\qty(\frac{\mathbb{I}-Z}{2}) = \frac{\mathbb{I} - 3Z}{2}\,,
    \quad\quad
    \hat{\rho}_3 = \mathcal{M}^{-1}\qty(\frac{\mathbb{I}-Z}{2}) = \frac{\mathbb{I} - 3Z}{2}\,,
    \quad\quad 
    \hat{\rho}_5 = \mathcal{M}^{-1}\qty(\frac{\mathbb{I}-Y}{2}) = \frac{\mathbb{I} - 3Y}{2}\,.
\end{aligned}
\end{equation}
Equations~\eqref{eq:app-classical-shadows} essentially tell that whenever the state $\rho$ is measured along the Pauli basis $P$, then its post-measurement classical approximation is defined as $\hat{\rho} = (\mathbb{I} + P)/2$ if eigenstate $+1$ was measured, and $\hat{\rho} = (\mathbb{I} - P)/2$ otherwise. With these classical snapshots at hand, one can then predict the expectation value on any observable $O$ computing the reconstruction coefficients $o_s = \Tr[O \hat{\rho}_s]$, and averaging over the different shots. 

\subsection{Estimation in the formalism of measurement frames and IC-POVMs}
We can now move our attention to the description of the same measurement process in the more general formalism of measurement frames and informationally-complete IC-POVMs. Far from being a complete introduction of frame theory, in this section we will only use well-known results of frame theory, and refer the interested reader to, e.g.~\cite{CasazzaFiniteFrames2013,CasazzaIntroFiniteFrames2013, KrahmerSparsityDualFrames2013} for general theory of frames in linear algebra, and to, e.g.~\cite{ShadowTomographyDualInnocenti2023, ScottTightICPOVM2006, ZhouOCPOVM2014, OptimalProcessingDariano2007, DArianoICMeasurements2004, Perinotti2007optimalestimationensembleaverages, Fischer2024DualOptimization, Malmi2024EnhancedEstimation}, for applications to quantum tomography. 

Roughly, a \textit{frame} for a vector space is a collection of vectors that spans the space, and it can consist of linearly dependent vectors. Frames provide a generalization of a basis of a vector space to that of an \textit{overcomplete} basis, in which case, due to the redundant degrees of freedom there exist many valid decompositions of a vector in terms of the frame elements.

In the context of quantum tomography, an informationally complete IC-POVM $\Pi = \{\Pi_k\}_k$ constitutes a \textit{frame} for the space of linear operators $\mathcal{L}(\mathcal{H})$ on the Hilbert space $\mathcal{H}$. According to frame theory, then any operator $X \in \mathcal{L}(\mathcal{H})$ can be expanded in terms of the POVM effects as~\cite{DArianoICMeasurements2004, OptimalProcessingDariano2007}
\begin{equation}
    \label{eq:app-obs-decomp}
    O = \sum_k \Tr[D_k O]\, \Pi_k\,.
\end{equation}
where $D = \{D_k\}_k$ is a so-called \textit{dual frame} to the measurement frame $\Pi = \{\Pi_k\}_k$. Importantly, whenever the POVM $\Pi$ consists of linearly dependent operators, that is, it is an \textit{overcomplete} (OC) POVM, then the dual frame is not unique, but infinitely many choices are possible~\cite{OptimalProcessingDariano2007}. 

Since the operator reconstruction coefficients $\omega_k = \Tr[D_k O]$ are non-unique for OC-POVMs, one has the freedom of choosing those that satisfy a given criterion, for example minimizing the variance of the operator (observable) estimation. The formalism of frames thus makes it evident that there are additional degrees of freedom in the observable reconstruction formula~\eqref{eq:app-obs-decomp}, and that these can be leveraged to find a decomposition achieving a low-variance estimation. On the contrary, as we will now see, classical shadows only pick a specific set of reconstruction coefficients, namely those corresponding to \textit{canonical duals}. We also note that recently several techniques have been proposed to find those duals achieving a low-variance estimation~\cite{Fischer2024DualOptimization, Malmi2024EnhancedEstimation, CaprottiDualOptimisation}. However, as discussed in the main text, these can either only be applied to specific types of measurements, or cannot scale to large system sizes.

Given a frame $\{\Pi_k\}$, one defines the (canonical) \textit{frame (super)operator} as the linear map acting as
\begin{equation}
    \label{eq:app-frame-operator}
    F_{\text{can}}(\bm{\cdot}) \coloneqq \sum_k \frac{1}{\Tr[\Pi_k]}\Tr[\bm{\cdot}~\Pi_k] \Pi_k\,.
\end{equation}
Noticing that operators $\Pi_k$~\eqref{eq:app-oc-pauli-effects} and $U^\dagger\dyad{b}U$ in Eq.~\eqref{eq:app_explicit_meas_channel} are the same up to a normalization factor, one can readily see that frame (super)operator $F_{\text{can}}(\bm{\cdot})$ correspond to the measurement channel $\mathcal{M}(\bm{\cdot})$~\eqref{eq:app_explicit_meas_channel} in the classical shadow formalism.

Among all possible dual frames, the so-called \textit{canonical} dual frame plays a central role in frame theory, whose elements $\{D_k^{\text{can}}\}_k$ are defined as
\begin{equation}
    D_k^{\text{can}} = \frac{1}{\Tr[\Pi_k]}F_{\text{can}}^{-1}\qty(\Pi_k)\,.
\end{equation}
Again, comparing this expression against Eqs.~\eqref{eq:app-classical-shadows}, makes it evident that classical shadows are the canonical duals associated with the measurement effects. Note that for simplicity we hereby neglected some subtleties related to the different definitions of \textit{canonical duals} used in the context of quantum tomography (see~\cite{ZhouOCPOVM2014}) and in general frame theory for linear algebra, see~\cite{ShadowTomographyDualInnocenti2023} for more details.

We now show how the classical shadows and the canonical duals agree on the specific case of considering the OC-POVM given by the six Pauli eigenstates~\eqref{eq:app-oc-pauli-effects}. Instead of dealing with operators $\Pi_k$, is more convenient to use the vectorized notation $\kket{\Pi_k}$ introduced in Appendix~\ref{app:vectorisation}, in which case the frame operator and the canonical duals read
\begin{equation}
    \label{eq:app-frame-operator-vec}
    F_{\text{can}} = \sum_k \frac{1}{\Tr[\Pi_k]} \kket{\Pi_k}\!\bbra{\Pi_k},\, \quad \kket{D_k^{\text{can}}} = \frac{1}{\Tr[\Pi_k]} F^{-1}\kket{\Pi_k}\,.
\end{equation}
With the vectorized Pauli basis defined as $\kket{\mathbb{I}} = [1,0,0,0],\, \kket{X} = [0,1,0,0],\, \kket{Y} = [0,0,1,0],\, \kket{Z} = [0,0,0,1]$ and using~\eqref{eq:app-ptm-vec} one can check that the six Pauli effects~\eqref{eq:app-oc-pauli-effects} can be written in vector notation in the Pauli basis as
\begin{equation}
\begin{aligned}
    \kket{\Pi_0} &= \frac{1}{3\sqrt{2}}[+1, 0, 0, +1]\,,\quad \kket{\Pi_1} = \frac{1}{3\sqrt{2}}[+1, 0, 0, -1]\,,\quad
    \kket{\Pi_2} = \frac{1}{3\sqrt{2}}[+1, +1, 0, 0]\,,\\
    \kket{\Pi_3} &= \frac{1}{3\sqrt{2}}[+1, -1, 0, 0]\,,\quad
    \kket{\Pi_4} = \frac{1}{3\sqrt{2}}[+1, 0, +1, 0]\,,\quad
    \kket{\Pi_5} = \frac{1}{3\sqrt{2}}[+1, 0, -1, 0]\,,
\end{aligned}
\end{equation}
and thus, by explicit computation, the frame operator~\eqref{eq:app-frame-operator-vec} in Pauli Transfer Matrix (PTM) form reads
\begin{equation}
    F_{\text{can}} = \text{diag}\qty(1,\, \frac{1}{3},\, \frac{1}{3},\, \frac{1}{3}) \implies F^{-1}_{\text{can}} = \text{diag}\qty(1,\, 3,\, 3,\, 3)\,.
\end{equation}
Noticing that a single-qubit depolarizing channel $\mathcal{D}_p(\rho) = p\rho + (1-p)\mathbb{I}/2$ has PTM $D_p = \diag(1, p, p, p)$~\cite{ManginiDeconvolution2022}, one finds that the frame operator is a depolarizing channel with $p=1/3$. Finally, one can check that the canonical duals $\kket{D_{k}^{\text{can}}} = F^{-1}_{\text{can}}\kket{\Pi_k} / \Tr[\Pi_k]$ indeed match the classical shadows $\hat{\rho}_s$ computed in Eq.~\eqref{eq:app-classical-shadows}.

\section{Optimization details} 
\label{app:optimization_details}
The variational sweeping optimization of the MPS $\kket{\omega}$ greatly benefits from standard tensor network techniques like canonization~\cite{Schollwok2011DMRGMPS}. Indeed, we find that the use of canonization helps both in speeding up the convergence to the solution, and also to reduce numerical instabilities arising from ill-conditioned matrices in~\eqref{eq:local-cost-solution}. In all the numerical experiments reported below, we use the MPS in mixed canonical form while sweeping through the sites.

Additionally, numerical instabilities can be ameliorated by adding a regularization term in the local cost function~\eqref{eq:local-cost} (\eg Tikhonov regularization), or for example choosing instead a least-square solution to the linear system. Additionally, one could also consider using a numerical optimizer to minimize the local cost~\eqref{eq:local-cost-solution} in addition to using the explicit solution provided by Eq.~\eqref{eq:local-cost}. In our experiments, we found that the explicit solution is capable of quickly converging to a good solution in a couple of sweeps in the case of infinite statistics, with more sweeps needed for finite statistics scenarios in general. 

Whenever optimization is run for finite statistics (see Sec.~\ref{sec:finite-ghz}), that is when we use the experimental counts instead of quantum probabilities in~\eqref{eq:tn-cost-fun}, we found it useful to modify the local tensors according to the updated rule
\begin{equation}
    \label{eq:smooth-update}
    \bm{\omega}_k^{\text{new}} = \alpha\,\bm{\omega}_k^{\text{opt}} + (1-\alpha)\,\bm{\omega}_k^{\text{old}}\,,
\end{equation}
which is a convex combination of the explicit solution $\bm{\omega}_k^{\text{opt}}$ of the local system~\eqref{eq:local-cost-solution} and the current values of the tensor at that site, balanced by hyperparameter $\alpha \in [0,1]$. Such an update rule avoids big jumps in the cost function and makes the whole optimization process more continuous. This not only helps in avoiding local minima, but also makes it easier to stop optimization whenever \textit{overfitting} of the training data is detected.

\section{Convergence guarantees for a biased estimator}
\label{app:performance-guarantee}
In this section, we discuss some convergence guarantees for the case of using a biased estimator to infer a quantity of interest. In particular, in Sec.~\ref{app:sec_chebychev} we first show how to derive the Chebyshev-like bound for the sample mean reported in the main text in Eq.~\eqref{eq:performance-guarantee}. Then, in Sec.~\ref{app:sec_hoeffding}, we show how tighter bounds with Hoeffding-like performances can be obtained for the median-of-means estimator and also for the sample mean, under additional assumptions on the distribution of the random variables. We now start by summarizing the estimation setting and fixing the notation. 

Let $\omega$ denote the tensor network estimator obtained at the end of the penalty-regularized variance minimization procedure, as discussed in Sec.~\ref{sec:tn-estimation}. If the optimization is successful, the reconstruction coefficients $\{\omega_k\}_k$ provide a low-variance statistical estimator which approximate the target observable $O$ with small error, that is
\begin{equation}
\label{eqa:app_approx_obs}
    O_\omega = \sum_k \omega_k \Pi_k\,, \quad \text{such that} \quad \norm{O - O_{\omega}}_2 = \varepsilon \ll 1\,.
\end{equation}
where $\{\Pi_k\}_k$ are the effects of the chosen IC-POVM, see Sec~\ref{sec:ic-measurement}.
Importantly, the requirement that the operators are close in operator space $\norm{O - O_{\omega}}_2 \leq \varepsilon$ also implies that their expectation values are close on any quantum state $\rho$ since,
\begin{equation}
    \label{eq:app_bound_approx_expval}
    \abs{\expval{O_\omega} - \expval{O}} = \abs{\Tr\qty[\qty(O_\omega - O)\rho]} \leq \norm{O_\omega - O}_2 \norm{\rho}_2 \leq \varepsilon\,,
\end{equation}
where we have used first Hödler's inequality, and then the fact that the purity of a quantum state is always lower than one $\norm{\rho}_2 = \Tr[\rho^2] \leq 1$.

A measurement on a state $\rho$ using a POVM with effects $\{\Pi_k\}_k$ will yield outcome $\Pi_{k}$ with probability given by Born's rule $p_k = \Tr[\Pi_k \rho]$. According to the observable decomposition formula Eq.~\eqref{eqa:app_approx_obs}, to each measurement outcome we have an associated reconstruction coefficient $\omega_k$ that can be used to estimate the expectation value $\expval{O_\omega} = \Tr[O_\omega \rho]$. Considered as random variables,  the reconstruction coefficients are distributed according to probability distribution $\{p_k\}_k$ with expectation and variance
\begin{equation}
\label{eq:app_stats_prop_w}
\begin{aligned}
   \mathbb{E}[\omega] & \coloneqq \sum_{k} p_k \omega_k = \sum_k \Tr[\Pi_k \rho] \omega_k = \Tr[O_\omega \rho] = \expval{O_\omega}\,,\\
    \Var{\omega} & \coloneqq \mathbb{E}[\omega^2] - \mathbb{E}[\omega]^2 = \sum_{k} p_k \omega_k^2 - \qty(\sum_k p_k \omega_k)^2\,.
\end{aligned}
\end{equation}
Let $\overbar{\omega}$ denote the sample mean estimator obtained by averaging the reconstruction coefficients $\{\omega_k\}_k$ observed in an experiment with $S$ measurement shots,
\begin{equation}
    \label{eq:app-empirical-mean}
    \overbar{\omega} \coloneqq \frac{1}{S}\sum_{s=1}^S \omega_{k_s}\,,
\end{equation}
where $\omega_{k_s}$ is the reconstruction coefficient corresponding to outcome $k_s$ obtained as outcome to the $s$-th measurement shot. By Eqs.~\eqref{eq:app_stats_prop_w}, the empirical mean $\overbar{\omega}$ provides an unbiased estimator to the observable $O_\omega$, that is
\begin{equation}
\label{eq:app_sample_means_props}
\begin{aligned}
    &\E{\overbar{\omega}} = \frac{1}{S}\sum_{s=1}^S \E{\omega_{k_s}} = \expval{O_\omega} \\
    &\Var{\overbar{\omega}} = \frac{1}{S^2} \sum_{s=1}^S \Var{\omega_{k_s}} = \frac{1}{S}\Var{\omega}\,,
\end{aligned}
\end{equation} 
where the random variables $\omega_{k_s}$ are statistically independent because they come from independent measurement shots. 

Keep in mind that our goal is to predict the expectation value of the true observable $\Tr[O\rho] = \expval{O}$ with good accuracy. However, we only have access to a $\varepsilon$-close approximation $O_\omega$ of the true observable $O$~\eqref{eqa:app_approx_obs}, and so the random variables $\{\omega_k\}_k$ will at worst provide a $\varepsilon$-biased estimation of the true mean, see Eq.~\eqref{eq:app_bound_approx_expval}. We can now proceed by showing how bias-dependent convergence guarantees can be straightforwardly derived also for biased estimators.

\subsection{Chebyshev-like performances for the sample mean}
\label{app:sec_chebychev}
One can quantify the power of the empirical mean estimator $\overbar{\omega}$~\eqref{eq:app-empirical-mean} to predict the true expectation value $\expval{O}$ by studying the probability that the two are far from each other, namely $\text{Pr}\qty(\abs{\overbar{\omega} - \expval{O}} > \delta)$. Such probability can be bounded from above as in Chebyshev's inequality as follows
\begin{equation}
    \label{eq:guarantee-proof}
    \begin{aligned}
        \text{Pr}\qty(\abs{\overbar{\omega} - \expval{O}} > \delta) & \leq \frac{\mathbb{E}\qty[\qty(\overbar{\omega} - \expval{O})^2]}{\delta^2} = \frac{\mathbb{E}\qty[(\overbar{\omega} - \expval{O_\omega} + \expval{O_\omega} - \expval{O})^2]}{\delta^2}\\
        & = \frac{1}{\delta^2}\qty(\underbrace{\E{(\overbar{\omega} - \expval{O_\omega})^2}}_{= \Var{\overbar{\omega}}} + 2\qty(\expval{O_\omega} - \expval{O})\underbrace{\E{\overbar{\omega} - \expval{O_\omega}}}_{=\E{\overbar{\omega}} - \expval{O_\omega} = 0} + \underbrace{\qty(\expval{O_\omega} - \expval{O})^2}_{\leq \varepsilon^2}) \\
        & \leq \frac{\Var{\omega}}{\delta^2 S} + \frac{\varepsilon^2}{\delta^2}\,,
    \end{aligned}
\end{equation}
where in the first line we used Markov's inequality $\text{Pr}(\abs{x}>a) \leq \E{x^2} / a^2$, and the second line we used Eqs.~\eqref{eq:app_sample_means_props}. Essentially, note that the formula above can be seen as a bias-variance decomposition of the expected mean-squared error of an estimator.

The bound in Eq.~\eqref{eq:guarantee-proof} comprises two terms. The first one is related to the statistical fluctuation of the reconstruction coefficients $\Var{\omega}$, which we assume is small because it was minimized during training (see Sec.~\ref{sec:tn-estimation}), and it can be further decreased by using a larger sample size $S$. The second one instead is independent of the number of shots, but takes into account the fact that the tensor estimator provides only a $\varepsilon$-close approximation of the true observable. 

Thus, assuming that the penalty-regularized variance minimization procedure is successful, that is we obtain a tensor estimator $\omega$ with low statistical error $\Var{\omega}$ and low reconstruction error $\norm{O_\omega -O}_2 = \varepsilon \ll 1$, then Eq.~\eqref{eq:guarantee-proof} guarantees that the estimated value will be close to the true expectation value.

\subsection{Improved Hoeffding-like convergence bounds}
\label{app:sec_hoeffding}
In addition to the Chebyshev-like concentration bound discussed above, tighter convergence bounds with Hoeffding-like performances could also be derived for the sample mean estimator $\overbar{\omega}$, under the additional assumption that its distribution is not heavy-tailed. 

First, note that the distance between the sample mean $\overbar{\omega}$ and the true expectation value $\expval{O}$ is upper bounded 
\begin{equation}
\label{eq:app_start_hoeff}
    \abs{\overbar{\omega} - \expval{O}} = \abs{\overbar{\omega} - \expval{O_\omega} + \expval{O_\omega} - \expval{O}} \leq \abs{\overbar{\omega} - \expval{O_\omega}} + \abs{ \expval{O_\omega} - \expval{O}} \leq \abs{\overbar{\omega} - \expval{O_\omega}} + \varepsilon\,,
\end{equation}
where the last inequality comes from~\eqref{eq:app_bound_approx_expval}. Considered as two random variables $x = \abs{\overbar{\omega} - \mu}$ and $y = \abs{\overbar{\omega} - \expval{O_\omega}} + \varepsilon$ both depending on the random variable $\overbar{\omega}$, the inequality $x \leq y$ or alternatively $\text{Pr}(x \leq y) = 1$, implies $\text{Pr}(x \geq \delta) \leq \text{Pr}(y \geq \delta)$ (see Theorem 1.A.1 in~\cite{Shaked2007Stochastic}). Thus, we can also write
\begin{equation}
    \label{eq:app_alternative_performance_proof}
    \text{Pr}(\abs{\overbar{\omega} - \expval{O}} \geq \delta) \leq \text{Pr}(\abs{\overbar{\omega} - \expval{O_\omega}} + \varepsilon \geq \delta) = \text{Pr}(\abs{\overbar{\omega} - \expval{O_\omega}} \geq \delta - \varepsilon)\,,
\end{equation}
which is meaningful as long as the reconstruction error $\varepsilon$ is smaller than the desired accuracy $\delta - \varepsilon \geq 0$. Equation~\eqref{eq:app_alternative_performance_proof} can then be used as the starting point to derive tighter convergence bounds for the sample mean estimator $\overbar{\omega}$ using standard concentration arguments based, for example, on Hoeffding's or Bernstein's inequalities~\cite{BoucheronConcentrationInequalitiesBook2013}, as done commonly in the classical shadow literature~\cite{HuangShadows2020, ShadowTomographyDualInnocenti2023, FermionicShadowTomography2021}. 

\subsubsection{Hoeffding-like performance using the sample mean}
In order to derive a tighter concentration bound for the sample mean, we have to add the additional assumption that the random variables $\{\omega_k\}_k$ take values in a restricted interval. In this case, one can then apply the well-known Hoeffding's inequality, which we report here for completeness.
\begin{theorem}[Hoeffding's inequality, see, e.g, Theorem 2.8 in~\cite{BoucheronConcentrationInequalitiesBook2013}, Theorem 8 in~\cite{KlieschQuantumCertification2021}]
\label{thm:hoeffding}
Let $X_1, \ldots, X_n$ be independent bounded random variables with $a \leq X_i \leq b$ almost surely $\forall i=1,\ldots,n$. Let $\overbar{X} =\qty(\sum_{i=1}^n X_i)/n$ denote their sample mean, then for all $t>0$ it holds
\begin{equation}
\begin{aligned}
    &\textup{Pr}\qty(\overbar{X} - \mathbb{E}\qty[\overbar{X}] \geq t) \leq \exp(-\frac{2 n t^2}{(b-a)^2})\,,\\
    &\textup{Pr}\qty(\overbar{X} - \mathbb{E}\qty[\overbar{X}] \leq - t) \leq \exp(-\frac{2 n t^2}{(b-a)^2})\,,\\
    &\textup{Pr}\qty(\abs{\overbar{X} - \mathbb{E}\qty[\overbar{X}]} \geq t) \leq 2\exp(-\frac{2 n t^2}{(b-a)^2})\,.
\end{aligned}
\end{equation}
\end{theorem}
Thus, assuming that the estimator coefficients $\{\omega_k\}_k$ distributed according to probabilities $\{p_k\}_k$, are \textit{bounded} random variables with $a \leq \omega_k \leq b~\forall k$, then by Hoeffding's inequality it holds that
\begin{equation}
    \label{eq:app_hoeffding}
    \text{Pr}\qty(\abs{\overbar{\omega} - \expval{O_\omega}} \geq t) \leq 2 \exp(-\frac{2 S t^2}{(b-a)^2}) \quad \quad \text{if}\quad a \leq \omega_k \leq b\quad\forall k\,,
\end{equation}
and by plugging this in Eq.~\eqref{eq:app_alternative_performance_proof} one obtains
\begin{equation}
    \text{Pr}(\abs{\overbar{\omega} - \expval{O}} \geq \delta) \leq \text{Pr}(\abs{\overbar{\omega} - \expval{O_\omega}} \geq \delta - \varepsilon) \leq 2 \exp(-\frac{2S(\delta - \varepsilon)^2}{(b-a)^2}) \quad\quad \,,
\end{equation}
which bounds the probability that the $\varepsilon$-biased estimator $\overbar{\omega}$ deviates from the desired true mean value $\mu = \expval{O}$.

Similar bounds can be derived under the less stringent assumption that the random variables $\omega_i$ are sub-Gaussian (instead of bounded), or improved taking into account the variance of the estimator using Bernstein's inequality, as proposed in~\cite{FermionicShadowTomography2021} in the context of classical shadows for fermionic systems. 

Importantly, as recently noticed by~\cite{AcharyaShadowTomographyPOVM2021, ShadowTomographyDualInnocenti2023, ThriftyShadowEstimation2023}, Hoeffding-like performances are achievable already with the sample mean estimator whenever the underlying distribution of the coefficients is well-behaved, that is, the observable reconstruction coefficients are bounded by a constant which does not scale exponentially with the system size (e.g. when estimating local observables). In these cases, the medians-of-means estimator originally proposed in~\cite{HuangShadows2020} can be substituted with the sample mean without loss of convergence guarantees. 

While it is possible to check this condition in the standard shadows protocol in which the reconstruction coefficients $\omega_k^{\text{can}} = \Tr[O D_k^{\text{can}}]$ have an explicit form in terms of canonical duals (or classical shadows) $D_k^{\text{can}}$~\cite{ShadowTomographyDualInnocenti2023}, this is generally not the case in our tensor estimator procedure, since the final tensor estimator is the result of a heuristic optimization procedure, and thus it is not possible to have an \textit{a priori} guarantee that the optimized reconstruction coefficients will lie on a restricted interval. 

However, as stressed in the main text in Sec.~\ref{sec:final-remarks}, note that our tensor estimator encompasses also the canonical estimator, and improves on it by providing a lower estimator variance (possibly at the cost of slightly increasing the range of the reconstruction coefficients). Thus, we expect the tensor estimator to lie approximately in the same range of the canonical estimator, hence to have rigorous convergence guarantees in the same settings of required by the canonical estimator.

\subsubsection{Hoeffding-like performance using the median-of-means estimator}
As discussed above, Hoeffding-like convergence guarantees are achievable with the empirical mean estimator whenever the reconstruction coefficients $\omega_k$ are bounded (or more generally, sub-Gaussian). In those cases in which such a condition cannot be met, then one can resort to the median-of-means trick, as originally proposed in the context of classical shadows in~\cite{Hu2021ShallowShadow}. Roughly, the median-of-means is an estimator that can achieve Hoeffding-like concentration guarantees also for heavy-tailed (not bounded) distribution, under the mild assumption that the random variables have finite variance~\cite{LugosiMeanEstimationReview2019, ShortnoteMom}.

While in all the analyses performed in this work we only deal with the sample mean estimator~\eqref{eq:app_sample_means_props}, for the sake of completeness we here show how one could derive a tight convergence bound for the median-of-means also in the case of a biased estimation process, as in our case. 

First of all, using Chebyshev's inequality in Eq.~\eqref{eq:app_alternative_performance_proof}, we obtain
\begin{equation}
   \text{Pr}(\abs{\overbar{\omega} - \expval{O}} \geq \delta) \leq \text{Pr}(\abs{\overbar{\omega} - \expval{O_\omega}} \geq \delta - \varepsilon) \leq \frac{\Var{\overbar{\omega}}}{(\delta-\epsilon)^2}\,, \quad\text{for}\quad \delta-\varepsilon \geq 0\,.
   \end{equation}
Setting $t = \Var{\overbar{\omega}}/(\delta-\epsilon)^2$ and hence $\delta = \sqrt{\Var{\overbar{\omega}}/t} + \varepsilon$, it means that with probability of at least $1 - t$ it holds
\begin{equation}
    \abs{\overbar{\omega} - \expval{O}} \leq \sqrt{\frac{\Var{\overbar{{\omega}}}}{t}} + \varepsilon\,.
\end{equation}
Specifically, setting $t=1/4$ and using that $\Var{\overbar{\omega}} = \Var{\omega} / S$~\eqref{eq:app_sample_means_props}, we have that with probability of at least 3/4 the following holds
\begin{equation}
    \label{eq:means_bound}
    \abs{\overbar{\omega} - \expval{O}} \leq \sqrt{\frac{4\Var{{\omega}}}{S}} + \varepsilon\,.
\end{equation}
We are now ready to prove the following concentration theorem for a biased estimator, which follows from a slight adaptation of the proof provided in Theorem 2 in~\cite{LugosiMeanEstimationReview2019}.

\begin{theorem}[adapted from Theorem 2 in~\cite{LugosiMeanEstimationReview2019}, see also Theorem 9 in~\cite{KlieschQuantumCertification2021}, Proposition 1 in~\cite{ShortnoteMom}]

Let $\omega_1, \ldots, \omega_N$ be independent random variables with mean $\mu_{\omega}$ and variance $\textup{Var}[\omega]$. Let the random variables $\{\omega_k\}_k$ be $\varepsilon$-biased with respect to a true mean of interest $\mu$, that is it holds that $\abs{\mu_\omega - \mu} \leq \varepsilon$. Let $N = S \cdot K$ with $S, K$ positive integers, then the median-of-means estimator $\omega_{\textup{mom}}$ obtained by clustering the total number of samples $N$ in $K$ clusters each of size $S$ defined as
\begin{equation}
    \omega_{\textup{mom}} \coloneqq \textup{median}(\overbar{\omega}_1, \ldots, \overbar{\omega}_K)\,, \quad\quad \overbar{\omega}_\ell = \frac{1}{S}\sum_{s = 1}^S\omega_{k_s}^{(\ell)}\,,
\end{equation}
satisfies
\begin{equation}
    \textup{Pr}\qty(\abs{\omega_{\textup{mom}} - \mu} \geq \sqrt{\frac{4 \textup{Var}[\omega]}{S}} + \varepsilon) \leq e^{-K/8}\,.
\end{equation}
In particular, for any $\delta \leq \varepsilon = (0,1)$, if $K = \lceil{8\ln(1/\delta)}\rceil$ then with probability of at least $1-\delta$,
\begin{equation}
    \label{eq:app_mom_hoeff}
    \abs{\omega_{\textup{mom}} - \mu} \leq \sqrt{\frac{32\textup{Var}[\omega]\log(1/\delta)}{N}} + \varepsilon\,.
\end{equation}
\end{theorem}

\begin{proof}
By Chebyshev's inequality~\eqref{eq:means_bound}, we have that, with probability at least 3/4, for each mean $\overbar{\omega}_\ell$ it holds
\begin{equation}
    \label{eq:means_bound_thm}
    \abs{\overbar{\omega}_\ell - \mu} \leq \sqrt{\frac{4\Var{{\omega}}}{S}} + \varepsilon\,, \quad \forall \ell =1, \ldots, K\,.
\end{equation}
For the median to be far from the true mean $\abs{\omega_{\text{mom}} - \mu} \geq \sqrt{4\Var{{\omega}}/S} + \varepsilon$ it must be that at least $K/2$ of the means are such that $\abs{\overbar{\omega}_{\ell} - \mu} \geq \sqrt{4\Var{{\omega}}/S} + \varepsilon$. Define the Bernoulli random variables
\begin{equation}
    z_\ell = 
    \begin{cases}
    1 \quad \text{if} \quad \abs{\overbar{\omega}_{\ell} - \mu} \geq \sqrt{4\Var{{\omega}}/S} + \varepsilon \\
    0 \quad \text{otherwise}
    \end{cases}\,
\end{equation}
for which, according to Eq.~\eqref{eq:means_bound_thm}, it holds $\Pr(z_\ell = 1) \leq 1/4$ and $\Pr(z_\ell = 0) \geq 3 / 4$. Consider the worst-case scenario, where the probability of large deviations is the largest, namely $\Pr(z_\ell = 1) = 1/4$ and $\Pr(z_\ell = 0) = 3 / 4$. The probability that at least $K/2$ of the means $\overbar{\omega}_\ell$ are far from the true mean $\mu$ can then be expressed in terms of the binomial random variable $\smash{z = \sum_{\ell=1}^K z_\ell \sim \text{Bin}(K, 1/4)}$, which counts how many means $\overbar{\omega}_\ell$ are far from the true mean $\mu$ by the given threshold. With this notation, we can then write,
\begin{align}
    \text{Pr} \qty(\abs{\omega_{\text{mom}} - \mu} \geq \sqrt{\frac{4\Var{{\omega}}}{S}} + \varepsilon) & \leq \text{Pr}\qty(\sum_{\ell=1}^K z_\ell \geq \frac{K}{2}) \\
    (\text{subtract }\mathbb{E}[z] = K/4 \text{ from both sides}) \quad\quad\quad & = \text{Pr}\qty(z - \mathbb{E}[z] \geq \frac{K}{4}) \\
    (\text{one-sided Hoeffding's inequality~\ref{thm:hoeffding}}) \quad\quad\quad & \leq e^{-K / 8}
\end{align}
Thus, if $K = \lceil{8\log(1/\delta)}\rceil$ and using $S = N/K$, then with probability of at least $1-t$ it holds
\begin{equation}
    \abs{\omega_{\text{mom}} - \mu} \leq \sqrt{\frac{4\Var{{\omega}}}{S}} + \varepsilon = \sqrt{\frac{32\Var{\omega}\log(1/t)}{N}} + \varepsilon
\end{equation}
\end{proof}

Summarizing, we have shown how one can generalize the proof for the median-of-means estimator also in which one has access only to a biased estimator of the quantity of interest. As one would expect, in this case the bound~\eqref{eq:app_mom_hoeff} guarantees that the biased median-of-means will soon converge to the true mean but within a constant offset that depends on the bias.

\section{Additional numerical results for finite statistics}
In this section, we report the full simulation results used to generate the plot in Fig-\ref{fig:finite_stats_ghz} in the main text. 

In the figure we show the optimization process of TN-ICE trained on datasets of different sizes $S = \{10^3, 10^4, 10^5, 10^6\}$ (train set) for the GHZ state on $n=6$ qubits and observable $O = X^{\otimes n} - Y^{\otimes n}$. During training, we monitor the performances of the estimator on an independent dataset of the same size (test set) to prevent overfitting of the training set. As visible from the plot, this happens when the training data is scarce and hence not representative of the underlying distribution. We refer to the main text for more comments on the results and overfitting. 

\label{app:additional-results}
\begin{figure}[ht]
    \centering
    \includegraphics[width = 0.85\textwidth]{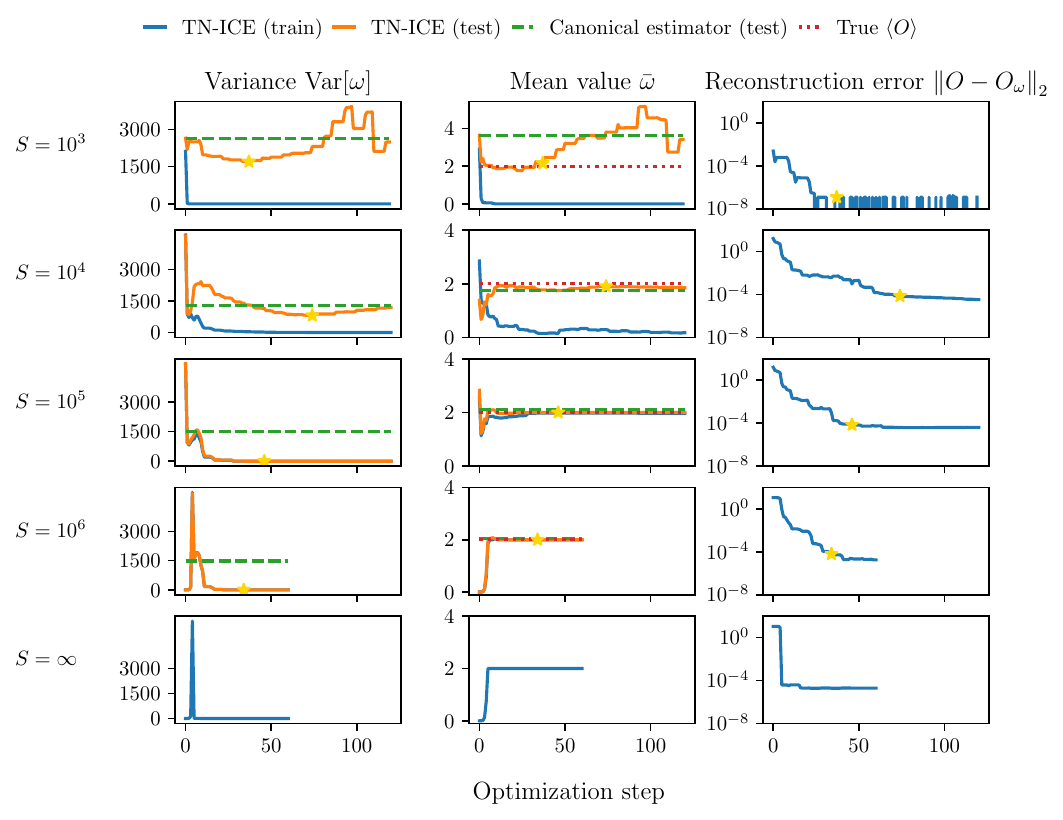}
    \caption{Full optimization runs for the data points reported in Fig.~\ref{fig:finite_stats_ghz} in the main text. For each size $S$, the tensor estimator is trained on a training dataset on size $S$ and its performances checked against an additional test set again of size $S$. The points marked with a yellow star are those reported in Fig.~\ref{fig:finite_stats_ghz} in the main text, and correspond to the points of minimum of the test variance and small reconstruction error. In addition to the finite statistics case, we here report also the results for training with exact probabilities, corresponding to infinite measurement budget $S=\infty$.}
    \label{fig:full-sim-ghz}
\end{figure}

\end{document}